\documentclass[twocolumn,showpacs,aps,amsmath,amssymb,superscriptaddress,prb]{revtex4-1}
\usepackage{amsfonts}
\usepackage{amsmath}
\usepackage{bm}
\usepackage{graphicx}
\usepackage{overpic}
\usepackage{color}
\usepackage{multirow}
\usepackage[normalem]{ulem}
\newcommand{\Sec}[1]{Sec.\,\ref{#1}}
\newcommand{\App}[1]{Appendix\,\ref{#1}}
\newcommand{\nl}{\nonumber \\}
\newcommand{\be}{\begin{equation}}
\newcommand{\ee}{\end{equation}}
\newcommand{\bea}{\begin{eqnarray}}
\newcommand{\eea}{\end{eqnarray}}
\newcommand{\Fig}[1]{Fig.~\ref{#1}}
\newcommand{\Figure}[1]{Figure~\ref{#1}}

\newcommand{\Eq}[1]{Eq.\,(\ref{#1})}
\newcommand{\Eqs}[1]{Eqs.\,(\ref{#1})}

\newcommand{\la}{\langle}
\newcommand{\ra}{\rangle}

\newcommand{\w}{\omega}
\newcommand{\ep}{\epsilon}
\newcommand{\D}{\Delta}

\newcommand{\TMPC}{T^{\ast,{\rm MPC}}}
\newcommand{\TZCC}{T^{\ast,{\rm ZCC}}}
\newcommand{\TiLMPC}{T_i^{\ast,{\rm LMPC}}}

\newcommand{\rhot}{\rho_{_{\rm T}}}
\newcommand{\HT}{\hat{H}_{_{\rm T}}}
\newcommand{\Hs}{\hat{H}_{_{\rm S}}}
\newcommand{\hH}{\hat{H}}
\newcommand{\trs}{{\rm tr}_{_{\rm S}}}
\newcommand{\trb}{{\rm tr}_{_{\rm B}}}
\newcommand{\trt}{{\rm tr}_{_{\rm T}}}

\begin{document}

\title{Effect of quantum resonances on local temperature in nonequilibrium open systems}

\author{Xiangzhong~Zeng}

\affiliation{Hefei National Laboratory for Physical Sciences at the
Microscale and Synergetic Innovation Center of Quantum Information and Quantum Physics,
University of Science and Technology of China, Hefei, Anhui 230026, China}

\author{Lyuzhou~Ye} 

\affiliation{Hefei National Laboratory for Physical Sciences at the
Microscale and Synergetic Innovation Center of Quantum Information and Quantum Physics,
University of Science and Technology of China, Hefei, Anhui 230026, China}

\author{Daochi~Zhang}

\affiliation{Hefei National Laboratory for Physical Sciences at the
Microscale and Synergetic Innovation Center of Quantum Information and Quantum Physics,
University of Science and Technology of China, Hefei, Anhui 230026, China}

\author{Rui-Xue~Xu}

\affiliation{Hefei National Laboratory for Physical Sciences at the
Microscale and Synergetic Innovation Center of Quantum Information and Quantum Physics,
University of Science and Technology of China, Hefei, Anhui 230026, China}

\author{Xiao~Zheng} \email{xz58@ustc.edu.cn}

\affiliation{Hefei National Laboratory for Physical Sciences at the
Microscale and Synergetic Innovation Center of Quantum Information and Quantum Physics,
University of Science and Technology of China, Hefei, Anhui 230026, China}

\author{Massimiliano Di Ventra} \email{diventra@physics.ucsd.edu}
\affiliation{Department of Physics, University of California, San Diego, La Jolla, California 92093, USA}

\date{Submitted on July~30, 2020; revised on October~22, 2020,January; accepted January~25, 2021; published February~4, 2021}

\begin{abstract}
  Measuring local temperatures of open systems out of equilibrium is emerging as a novel approach to
  study the local thermodynamic properties of nanosystems.
  An operational protocol 
  has been proposed to determine the local temperature by coupling a probe to the system
  and then minimizing the perturbation to a certain local observable of the probed system.
  In this paper, we first show that such a local temperature is unique for a single quantum impurity and the given local observable. We then extend this protocol to open systems consisting of multiple quantum impurities
  by proposing a local minimal perturbation condition (LMPC).
  The influence of quantum resonances on the local temperature is elucidated by both analytic and numerical results.
  In particular, we demonstrate that quantum resonances may give rise to strong oscillations
  of the local temperature along a multi-impurity chain under a thermal bias.
  \end{abstract}

\maketitle

\section{INTRODUCTION} \label{Inc}

Local temperatures of systems out of equilibrium\cite{Zhang2019local} are of fundamental importance
in many sub-fields of modern science, including physics,\cite{Hof09779,menges2013thermal,Men1610874,Inu184304} chemistry\cite{Lee13209,Cui18122,Nov19016806} and biology.\cite{Zei04871,Kuc1354,He1626737}
With the development of high-resolution thermometric techniques,\cite{Sadat2012High,mecklenburg2015Nanoscale,Men1610874}
the measurement of local temperature distributions in nonequilibrium nanoscopic systems has been realized,
such as in graphene-metal contacts,\cite{Gro11287} aluminum nanowires,\cite{mecklenburg2015Nanoscale} and two-dimensional metallic films.\cite{Gurrum2005Scanning}

A nonequilibrium system under an external driving source, such as a bias voltage or a thermal gradient,
often possesses a local temperature somewhat higher than the background temperature.\cite{di2008electrical}
Such a local heating effect usually originates from the electronic and phononic excitations occurring in the system
and has significant influence on some physical properties\cite{Liu2015Density}
and processes.\cite{Hua061240,huang2007local,Kim14203107,Thi15854,Ivo16014301,Idr18095901}

Theoretically, the concept of temperature has been extended from equilibrium systems to local subregions of nonequilibrium systems.\cite{scovil1959three,curzon1975efficiency,lieb1999physics,allahverdyan2001breakdown,
casas2003temperature,kieu2004second,bustamante2005nonequilibrium,horhammer2008information,bergfield2009many,
levy2012quantum,horodecki2013fundamental,skrzypczyk2014work,hardal2015superradiant,clos2016time,
puglisi2017temperature,marcantoni2017entropy,monsel2018autonomous,bialas2019quantum,Zhang2019local}
Ideally, the definition of local temperature should be universal
(so that it can be applied to as many nonequilibrium situations as possible),
unique (so that it yields one and only one value of temperature),
operationally feasible (so that it can be measured experimentally),
and has the correct asymptotic limit (so that it retrieves the thermodynamic temperature
as the system approaches an equilibrium state).\cite{Zhang2019local}

For instance, 
Engquist and Anderson have proposed a procedure for measuring the local temperature and local chemical potential of nonequilibrium systems.\cite{engquist1981definition}
In their definition, the local temperature and local chemical potential are measured, respectively, by a thermometer in thermal equilibrium with the system of interest and a potentiometer in electrical equilibrium with the system.
Later, Bergfield and Stafford \emph{et al.} pointed out that the above definition implicitly ignores thermoelectric effects and is nonunique.\cite{bergfield2013probing,Bergfield2014Thermoelectric,shastry2020scanning}
They have proposed a definition in which the local temperature and local chemical potential of a nonequilibrium system are determined simultaneously.\cite{bergfield2013probing,bergfield2015tunable,shastry2016temperature,shastry2020scanning}
In their definition, a probe which plays the roles of both potentiometer and thermometer
is weakly coupled to the nonequilibrium system of interest.
By varying the temperature ($T_p$) and chemical potential ($\mu_p$) of the probe until
both the electric ($I_p$) and heat currents ($J_p$) flowing through the probe vanish,
the local temperature ($T^\ast$) and local chemical potential ($\mu^\ast$) of the system
are determined as $T^\ast=T_p$ and $\mu^\ast=\mu_p$, respectively.
Such a condition is referred to as the zero current condition (ZCC).\cite{Zhang2019local}

The ZCC-based definition has been used to investigate local temperature\cite{Meair2014Local,Bergfield2014Thermoelectric,Shastry2015Cold,Stafford2016Local,Stafford2017Local} and local electrochemical potential distribution\cite{bevan2014first,morr2016crossover,morr2017scanning} in nanosystems out of equilibrium.\cite{caso2010local,caso2011local,caso2012defining}
It was found that the local temperature ($T^{\ast,{\rm ZCC}}$) exhibits an oscillatory behavior in nanowires,\cite{caso2011local} conjugated organic molecules\cite{bergfield2013probing} and graphene sheets.\cite{bergfield2015tunable}
Such oscillations originate from the emergence of quantum coherence
as the size of the system reduces to be comparable to or even smaller than the mean free path of electrons.\cite{dubi2009thermoelectric}
Consequently, the classical Fourier's law is strongly violated.\cite{michel2003fourier,dhar2008heat,roy2008crossover,yang2010violation,Zhang2019local}
Although it has been demonstrated\cite{dubi2009thermoelectric,caso2010local,caso2011local,bergfield2013probing,Meair2014Local,bergfield2015tunable,Inu184304} that quantum coherence and quantum interference effects could be captured by $T^{\ast,{\rm ZCC}}$, 
it was also shown that $T^{\ast,{\rm ZCC}}$ of a quantum dot experiences little change as the dot is tuned
from an off-resonance region into a resonance region.\cite{ye2016thermodynamic}
Therefore, it remains unclear whether $T^{\ast,{\rm ZCC}}$ could reflect the emergence of sharp quantum resonances.

The ZCC-based definition has the advantage of yielding a unique value of local temperature for any nonequilibrium system,\cite{shastry2016temperature} and it is applicable even in situations where local equilibrium states do not exist.\cite{Meair2014Local}
However, its experimental realization is rather challenging because of the difficulty in measuring the heat current through
a nanosized sample without knowing its priori local temperature.\cite{Zhang2019local}
Recently, Shastry and co-workers have modified the ZCC-based protocol,\cite{shastry2020scanning}
and the new protocol does not require the measurement of heat current if the system obeys the Wiedemann-Franz law.
However, the Wiedemann-Franz law is known to be violated by systems involving strong many-body interactions\cite{crossno2016observation} or low-energy quantum resonant states.\cite{Ye2014Thermopower}

Alternatively, an operational protocol to determine the local temperature has been proposed based on a minimal-perturbation condition (MPC),\cite{dubi2009thermoelectric,Ye2015local,ye2016thermodynamic}
with which the local temperature ($T^{\ast,{\rm MPC}}$) is determined by tuning $T_p$ and $\mu_p$ of a weakly coupled probe until the perturbation caused by the probe to the system gets minimized.

Ideally, if the perturbation caused by the probe to the system can be nullified, any dissipation between the system and the probe, particularly the energy and particle flows, would vanish. In such cases, the MPC can be deemed as a generalization of the zeroth law, and it naturally leads to the ZCC; see the formal proof in \App{app:proof}. Meanwhile, as a result of zero dissipation, any observable of the system preserves its intrinsic value as in the absence of the probe. 
However, in some occasions, such as when quantum resonances come into play,\cite{ye2016thermodynamic} the influence of the probe cannot be fully suppressed.  In the latter case, the MPC stands as an empirical principle which minimizes the dissipation between the system and the probe, while its thermodynamic meaning is less transparent. 
Because of the difficulty in measuring the heat current directly, in practice the {\it dissipation} is characterized by the particle current as well as the change of a certain system observable. The resulting MPC-based protocol has been applied to investigate local temperatures of strongly correlated quantum impurity systems.\cite{Ye2015local,ye2016thermodynamic}	

Despite the effectiveness and feasibility of the MPC-based definition, there are still issues that remain unclear.
Some of them are as follows:
(i) Does the MPC predict a unique value of $T^\ast$?
(ii) Why and how does $T^{\ast,{\rm MPC}}$ differ from $T^{\ast,{\rm ZCC}}$ in systems involving quantum resonances?
(iii) So far the application of the MPC has been restricted to systems containing a single impurity.
How do we extend the definition of $T^{\ast,{\rm MPC}}$ to multi-impurity systems?
(iv) Do quantum resonances lead to any discernible feature in the distribution of local temperatures along a quantum wire?

This work aims at elucidating the above issues through theoretical analysis and numerical calculations.
In particular, to address the last two questions, we propose a local minimal-perturbation condition (LMPC)
by imposing explicitly the nonequilibrium-equilibrium correspondence relation.
As a direct extension of the original MPC, the LMPC enables the determination of the local temperature
and local chemical potential of each impurity in a multi-impurity system out of equilibrium.

The remainder of this paper is organized as follows.
In \Sec{thsec2}, we revisit the MPC protocol and discuss how to reach a unique prediction of the local temperature of a single quantum
impurity.
In \Sec{thsec-3}, we propose the LMPC-based definition of local temperature. As a numerical example we
calculate the distribution of local temperatures along a quantum wire consisting of four impurities.
Concluding remarks are given in \Sec{thconc}.

\section{Effect of quantum resonances on local temperatures of single impurity systems} \label{thsec2}

\subsection{Quantum impurity systems} \label{thsec2B}

In the following, the Anderson impurity models (AIMs)\cite{anderson1961localized}
are adopted to describe the open systems.
The total Hamiltonian of the system plus environment is
\be
\hat H = \hat H_{\rm imp}+ \hat H_{\rm lead} + \hat H_{\rm coup}, \label{H-total}
\ee
where the three terms on the right-hand side (RHS) represent the Hamiltonian of the impurities,
the Hamiltonian of the leads which serve as the electron reservoirs and heat baths, and the Hamiltonian of
the impurity-lead couplings, respectively.

We consider first a single impurity described by
$\hat H_{\rm imp} = \epsilon_d \, \hat{n} + U \hat{n}_{\uparrow} \, \hat{n}_{\downarrow}$.
Here, $\hat{n} = \sum_s \hat{n}_s = \sum_s \hat a^\dag_s \, \hat a_s$, with
$\hat a^\dag_s$ ($\hat a_s$) creating (annihilating) an electron of spin $s$ on the impurity level $\ep_d$,
and $U$ is the Coulomb interaction energy between the spin-up and spin-down electrons.
$\hat H_{\rm lead}=\sum_{\alpha k s} \epsilon_{\alpha k} \, \hat{d}^\dag_{\alpha k s} \, \hat{d}_{\alpha k s}$ and $\hat H_{\rm coup}=\sum_{\alpha k s} t_{\alpha k} \, \hat{a}^\dag_s \, \hat{d}_{\alpha k s} + {\rm H.c.}$ describe the noninteracting leads and the impurity-lead coupling, respectively.
Here, $\hat{d}^\dag_{\alpha k s}(\hat{d}_{\alpha k s})$ creates (annihilates) a spin-$s$ electron on the $k$th orbital of the $\alpha$th lead, 
and $t_{\alpha k}$ is the coupling strength between the impurity level and the $k$th lead orbital.

To investigate the properties of the impurity, the hierarchical equations of motion (HEOM) approach\cite{Tanimura1989Time,jin2008exact,li2012hierarchical,Cui2019Highly,Zhang2020Hierarchical,Tanimura2020Numerically} is employed,
which takes the reduced density matrix of the impurity and a hierarchical set of auxiliary density operators as the basic variables.
%
The HEOM theory is, in principle, formally exact, and its numerical outcomes are guaranteed to be quantitatively accurate
if the results converge with respect to the truncation tier of the hierarchy ($L$).\cite{Ye16608,Han2018On}
For noninteracting impurities ($U=0$), the HEOM theory is formally equivalent to the nonequilibrium Green's function (NEGF) formalism,\cite{jin2008exact} and a low truncation tier of $L=2$ suffices to yield exact single-electron properties.
For interacting AIMs ($U \ne 0$), while the analysis with the NEGF method is difficult because the self-energies due to 
electron-electron interaction cannot be expressed analytically,
the HEOM approach can still yield accurate numerical results (usually with a truncation tier of $L > 2$).\cite{Zheng2009Numerical,zheng2013kondo,Wang2013Hierarchical,Ye2014Thermopower,Hou2015Improving,Ye2015local,ye2016thermodynamic,Wang2018Precise,li2020molecular}
%
%
%

In the framework of the HEOM, the influence of the noninteracting leads on the impurity is fully captured by the hybridization functions, $\Gamma_{\alpha}(\omega)\equiv \pi\sum_k|t_{\alpha k}|^2\delta(\omega-\epsilon_{\alpha k})$.
For numerical convenience, a Lorentzian form of
$\Gamma_{\alpha}(\omega)=\frac{\Delta_{\alpha} W_\alpha^2}{(\omega-\Omega_\alpha)^2+W_\alpha^2}$ is adopted,
where $\Delta_{\alpha}$ is the effective coupling strength between the impurity and the $\alpha$th lead,
and $\Omega_\alpha$ and $W_\alpha$ are the band center and width of the $\alpha$th lead, respectively.
We further set the band center at the chemical potential of the lead, i.e., $\Omega_\alpha = \mu_\alpha$.
The chemical potential of the equilibrium composite system is set to the zero energy, i.e., $\mu^{\rm eq} = 0$.

In the following, $\Delta_p$ (the subscript $p$ denotes the probe) is taken to be at least two orders of magnitude smaller
than all the other $\Delta_\alpha$, and further reducing its value does not influence the resulting $T^\ast$ and $\mu^\ast$.\cite{Ye2015local}
Hereafter, we adopt the atomic units $e=\hbar = k_{\rm B} \equiv 1$, and
$\Delta = \sum_{\alpha \neq p} \Delta_\alpha$ is taken as the unit of energy.

\subsection{Minimal-perturbation condition} \label{thsec2A}

The MPC-based protocol is generally applicable to {\it any} quantum impurity system regardless of the specific form of the system Hamiltonian. Here, we consider
the scenario that a single impurity is coupled to the left ($L$) and right ($R$) leads,
whose background temperatures (chemical potentials) are $T_L$ and $T_R$ ($\mu_L$ and $\mu_R$), respectively.
By coupling an external probe to the impurity, the local observable $O = \la \hat O \ra =  {\rm tr}(\hat{O}\rho)$
is subject to a perturbation of
\be
\delta O_p = O_p(T_p,\mu_p) - O_{\rm ref}. \label{m-3}
\ee
Here, $O_p(T_p,\mu_p)$ is the value of $O$ measured by setting the temperature and chemical potential of the coupled probe to $T_p$ and $\mu_p$, respectively. $O_{\rm ref}$ is the minimally perturbed value of $O$ which serves as reference for $O_p$.
For a single-impurity system, $O_{\rm ref}$ is determined by
\be
O_{\rm ref} = \zeta_L O_p(T_L,\mu_L)+ \zeta_R O_p(T_R,\mu_R), \label{Oref-1}
\ee
where the coefficients $\zeta_\alpha$ ($\alpha=L$ and $R$) are acquired as\cite{Ye2015local}
\be
  \zeta_\alpha = 1 - \left| \frac{I_p(T_\alpha, \mu_\alpha)}{I_p(T_L, \mu_L)-I_p(T_R, \mu_R)} \right|. \label{zeta-alpha-1}
\ee
Here, $I_p(T_p,\mu_p)$ is the electric current flowing into the probe with its temperature and chemical potential set to $T_p$ and $\mu_p$, respectively.
For AIMs, $\zeta_\alpha = \frac{\Delta_\alpha}{\Delta_L + \Delta_R}$.

The local temperature $T^\ast$ and local chemical potential $\mu^\ast$ of the impurity are determined by
\begin{equation} \label{def-Tloc-5}
\left\{
\begin{aligned}
 & {I_{p}(T^\ast, \mu^\ast) = 0}  \\
 & (T^\ast, \mu^\ast) = {\arg\min_{(T_p, \mu_p)}|\delta O_{p}(T_p, \mu_p)|}
\end{aligned}
\right. .
\end{equation}
In particular, if $\delta O_p= 0$ is achievable by tuning $T_p$ and $\mu_p$,
the MPC actually becomes the zero perturbation condition (ZPC).
The definition of \Eq{def-Tloc-5} does not involve the troublesome heat current,
and all the involving quantities (e.g. $I_p$ and $O_p$) can be measured directly in experiments.
Therefore, \Eq{def-Tloc-5} provides an operational protocol for the determination of $T^\ast$ and $\mu^\ast$.
In practice, such a protocol for the system under a bias voltage may be further simplified with a preset $\mu^\ast$ as follows,
\begin{equation} \label{def-Tloc-55}
\left\{
\begin{aligned}
& {\mu^\ast \approx \zeta_L \mu_L+ \zeta_R \mu_R} \\
& {T^\ast = \arg \min_{T_p}|\delta O_p(T_p,\mu^\ast)|}
\end{aligned}
\right. .
\end{equation}

For a convenient and accurate measurement of $T^\ast$, the local observable $O$ should vary sensitively with $T_p$.
In our previous works,\cite{Ye2015local,ye2016thermodynamic} the local magnetic and charge susceptibilities of the impurity,
$\chi^m  = \frac{\partial \langle \hat{m}_z \rangle}{\partial H_z}|_{H_z \rightarrow 0}$
and $\chi^c  = -\frac{\partial \langle \hat{n} \rangle}{\partial \epsilon_d}$, respectively,
have been chosen as the local observables.
Here, $\hat{m}_z = \frac{1}{2} g\,\mu_{_{\rm B}}(\hat{n}_\uparrow - \hat{n}_\downarrow)$ is the impurity magnetization operator,
with $H_z$ being the local magnetic field, $g$ the electron gyromagnetic ratio, and $\mu_{_{\rm B}}$ the Bohr magneton.
It has been shown that while $T^{\ast,{\rm MPC}}(\chi^m)$ and $T^{\ast,{\rm MPC}}(\chi^c)$ agree closely with each other in most cases,
they do exhibit small discrepancy in the near-resonance (NR) region.\cite{ye2016thermodynamic}

\begin{figure}[t]
\centering
\includegraphics[width=\columnwidth]{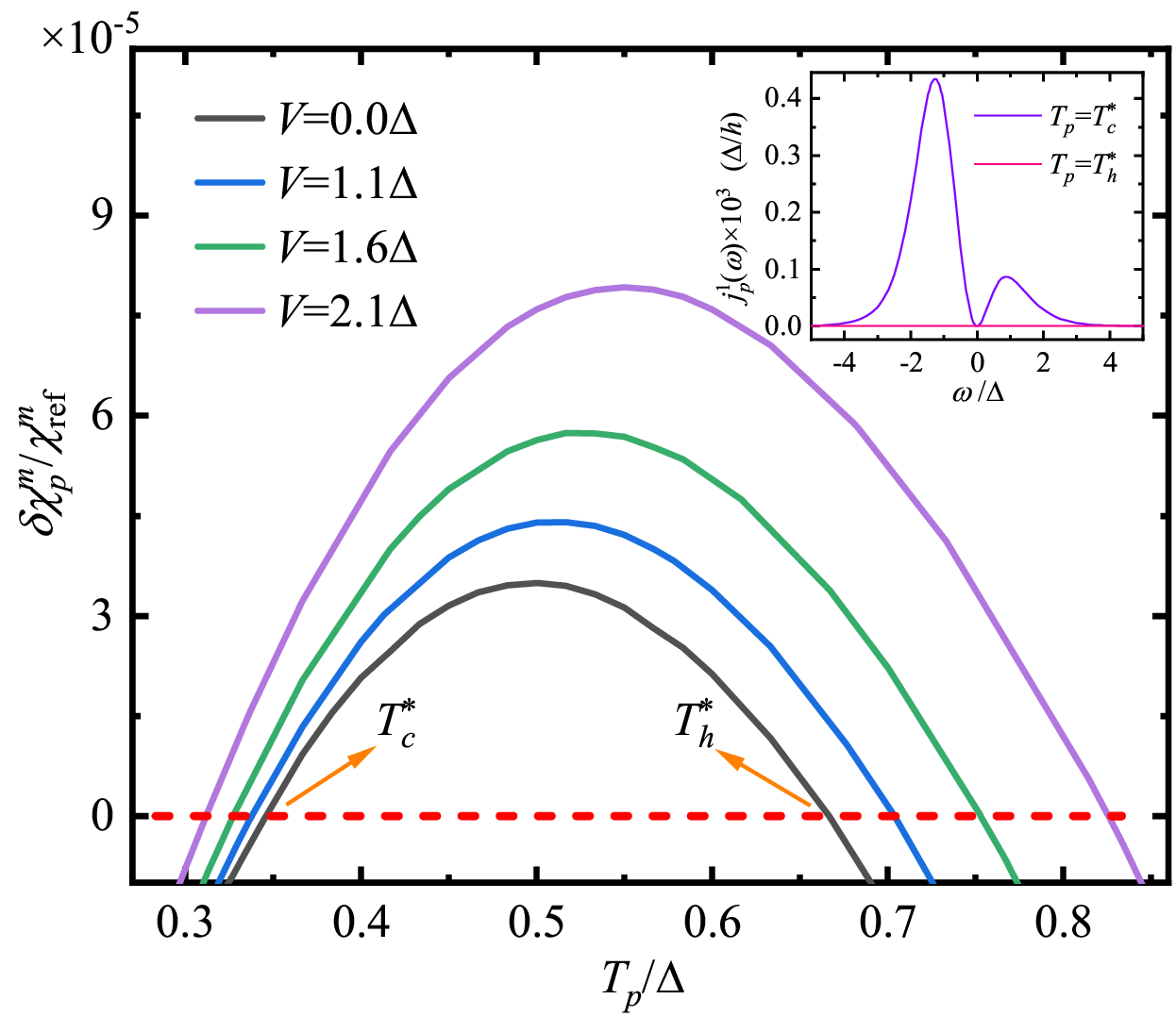}
\caption{$\delta \chi^m_p / \chi^m_{\rm ref}$ as a function of $T_p$ for a single-impurity
system under an antisymmetric bias voltage of $\mu_R = -\mu_L = V/2$.
$T^\ast_c$ and $T^\ast_h$ are two temperatures which satisfy the ZPC of $\delta \chi^m_p = 0$, as indicated by the horizontal line.
The energetic parameters of the system are (in units of $\Delta$):
$T_L = T_R = 0.67$, $\epsilon_d=-3.33$, $\Delta_L =\Delta_R = 0.5$, $U=2.5$, and $W_L = W_R = 6.67$.
The inset depicts the energy distribution of the heat current flowing into the probe [$j_p^1(\omega)$] at $T_p=T_c^\ast$ and $T_p=T_h^\ast$ under the zero bias, respectively. The detailed expression of $j_p^1(\omega)$ has been given in Ref.~[\onlinecite{Ye2015local}], and $J_p=\int j_p^1(\omega)d\omega$.} \label{figa1}
\end{figure}

In the following, we explore the uniqueness/nonuniqueness of the $\TMPC$.
First, we show that the MPC of \Eq{def-Tloc-5} or \Eq{def-Tloc-55} may give rise to multiple values of $T^\ast$ with a certain $O$.
Figure~\ref{figa1} depicts the relative perturbation of local magnetic susceptibility,
$\delta\chi^m_p/\chi_{\rm ref}^m$, as a function of $T_p$
for a single-impurity system under an antisymmetric bias voltage.
From the first line of \Eq{def-Tloc-55} we have $\mu^\ast = 0$ since $\Delta_L = \Delta_R$.
Meanwhile, it is intriguing to find that there are two temperatures that could satisfy the ZPC of $\delta\chi^m_p = 0$,
which are designated as $T^\ast_c$ and $T^\ast_h$ ($T^\ast_c < T^\ast_h$).
Thus, the local temperature $\TMPC$ appears to be nonunique.
However, it is important to note that, as the bias voltage is reduced the whole system should evolve towards an equilibrium state. In particular, in the limit of $V \rightarrow 0$,
$\TMPC$ should recover the thermodynamic temperature of the equilibrium system, i.e., $\TMPC = T_L = T_R$.
In particular, as indicated by the inset of \Fig{figa1}, the heat current through the probe ($J_p$) vanishes only at $T_p=T_h^\ast=T_L=T_R$, while it retains an appreciable value at $T_p=T_c^\ast$.
It is thus evident that only $T^\ast_h$ achieves the correct asymptotic limit under the zero bias and recovers the zeroth law.
Therefore, although the MPC of \Eq{def-Tloc-55} has multiple solutions, because it does not explicitly examine the heat current, the $\TMPC$ can still be uniquely determined for a given local observable $O$ by considering the asymptotic limit of the global equilibrium state.

\subsection{Effect of quantum resonances on local temperature}
\label{thsec3B}

\begin{figure}[t]
  \centering
  \includegraphics[width=\columnwidth]{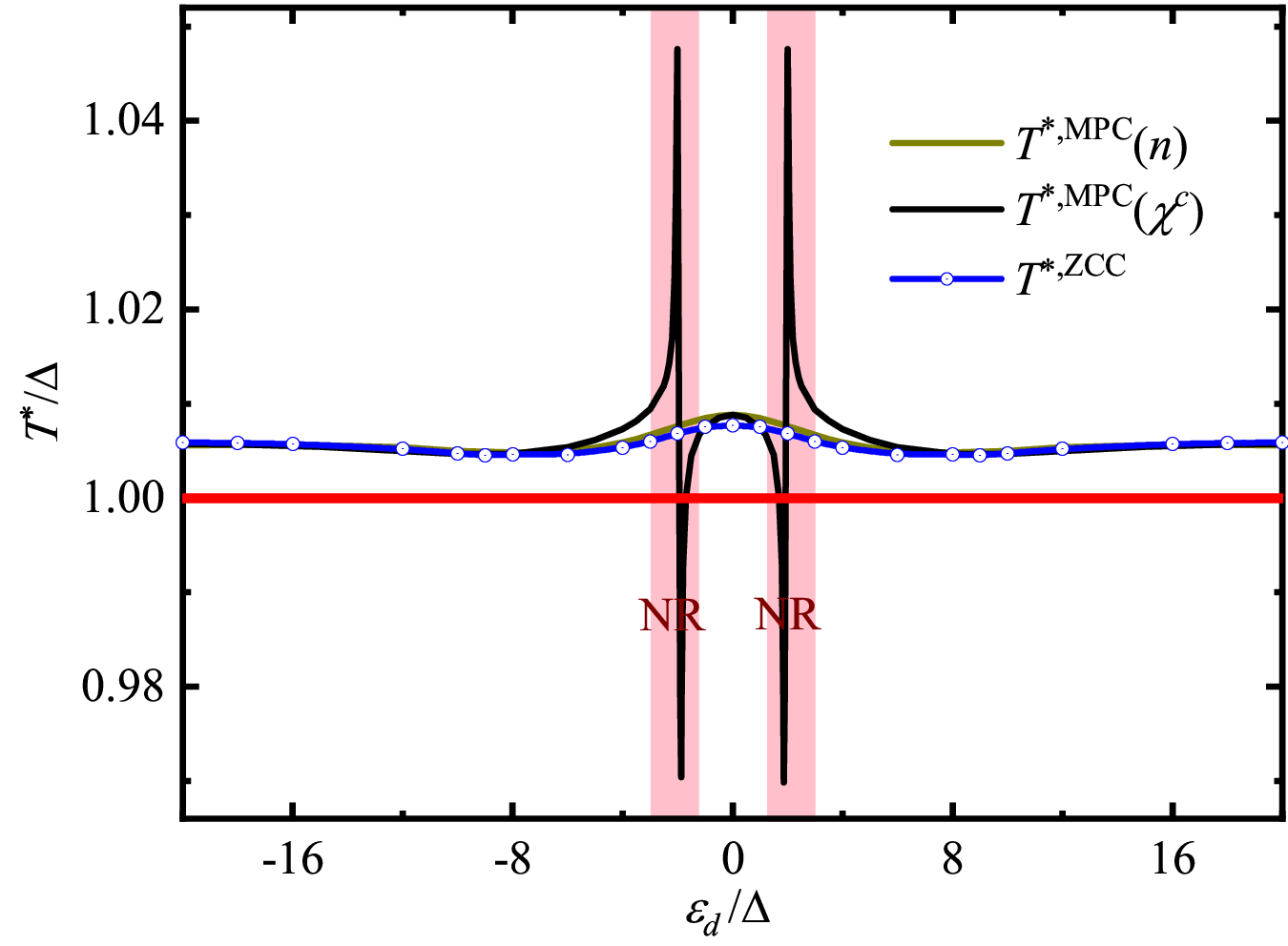}
  \caption{$\TMPC(n)$, $\TMPC(\chi^c)$, and $T^{\ast,{\rm ZCC}}$ versus $\epsilon_d$ for a noninteracting single-impurity system
  under an antisymmetric bias voltage of $\mu_R = -\mu_L = V/2 = 0.2\Delta$.
  The energetic parameters of the system are (in units of $\Delta$): $T_L = T_R = 1$, $U=0$,
  $\Delta_L =\Delta_R = 0.5$, and $W_L = W_R = 50$.
  The shaded areas represent the NR regions.
  } \label{figa2}
  \end{figure}

We then explore the uniqueness/nonuniqueness of $\TMPC$ associated with different local observables
in the off-resonance, near-resonance, and resonance regions.
To this end, we consider a noninteracting single-impurity system under an antisymmetric bias voltage.
Figure~\ref{figa2} depicts the variation of $\TMPC$ determined by \Eq{def-Tloc-55} with the change of $\ep_d$.
The displayed $\TMPC$ are associated with the electron occupation number on the impurity $n = \la \hat{n} \ra$ or
with the local charge susceptibility $\chi^c$. In both cases, we have $\mu^{\ast,{\rm MPC}} = 0$ since $\Delta_L = \Delta_R$.
For comparison, the $\TZCC$ versus $\ep_d$ are also shown in \Fig{figa2}.

From \Fig{figa2}, it is clear that $T^{\ast,{\rm ZCC}}$ and $T^{\ast,{\rm MPC}}(n)$ vary smoothly
and coincide closely with each other over the whole range of $\epsilon_d$.
In contrast, while the $T^{\ast,{\rm MPC}}(\chi^c)$ agree well with the other two local temperatures in the off-resonance regions ($|\ep_d|$ is far away from the chemical potentials of leads), they exhibit strong oscillations in the NR regions.
Such oscillations reflect the emergence of nonlocal excitations as a quantum resonant state begins to establish in the system.\cite{ye2016thermodynamic}
%

To understand the quantitative agreement between $T^{\ast,{\rm MPC}}(n)$ and $T^{\ast,{\rm ZCC}}$ in \Fig{figa2},
we carry out some theoretical analysis by using the NEGF method. 
In the wide-band limit ($W \rightarrow\infty$), the spin-$s$ component of the steady-state electric current flowing into the probe is
\begin{align}
I_{p s} &= -\frac{i}{\pi} \int d\omega\, \Gamma_p(\omega) \left\{G_s^<(\omega)+ 2if_{T_p,\mu_p}(\omega)\, {\rm Im}[G^r_s(\omega)] \right\} \nonumber \\
&= \frac{2\Delta_p\Delta}{\Delta+\Delta_p}\int d\omega \,
A_s(\omega)\, \big[\zeta_L f_{T_{L},\mu_{L}}(\omega) \nonumber \\
&\qquad \qquad \qquad +\zeta_R f_{T_{R},\mu_{R}}(\omega)-f_{T_{p},\mu_{p}(\omega)} \big], \label{zcc-1}
\end{align}
and the electron occupation number on the impurity is
\begin{align}
n &= \sum_{s} n_s = \sum_s \frac{1}{2\pi i}\int d\omega\, G_s^<(\omega)  \nonumber \\
&= \sum_\alpha \frac{\Delta_\alpha}{\Delta + \Delta_p} \int d\omega \, A(\omega)f_{T_{\alpha},\mu_{\alpha}}(\omega).
\end{align}
Here, $G_s^r(\omega)$ and $G_s^<(\omega)$ are the retarded and lesser single-electron Green's functions of the impurity, respectively;
 $A(\w) = \sum_s A_s(\omega)=-\frac{1}{\pi} \sum_s {\rm Im}[G_s^r(\omega)]$ is the spectral function of the impurity, and $f_{T_\alpha,\mu_\alpha}(\omega)$ is the Fermi distribution function.

The ZPC for the local observable $n = \la \hat{n} \ra$ is expressed as
\begin{align}
\delta n_p &= n_{p}(T_p,\mu_p)-n_{\rm ref} \nonumber \\
&= \frac{\Delta_p}{\Delta+\Delta_p}\int d\omega \, A(\omega)\, \big\{f_{T_{p},\mu_{p}}(\omega) \nonumber \\
& \qquad \qquad -[\zeta_L f_{T_{L},\mu_{L}}(\omega)+\zeta_R f_{T_{R},\mu_{R}}(\omega)] \big\}  \nl
 & = 0. \label{zpc-1}
\end{align}
By comparing \Eqs{zcc-1} and \eqref{zpc-1}, it is immediately recognized that the ZPC for the observable $n$
is exactly equivalent to the ZCC of $I_p = \sum_s I_{ps} = 0$.

On the other hand, unlike the presumed $\mu^{\ast,{\rm MPC}}(n)=0$, the ZCC also requires zero heat current, $J_p = 0$,
which often gives rise to a nonzero $\mu^{\ast,{\rm ZCC}}$.
Such a minor difference in $\mu^\ast$ in turn leads to the slightly different $T^\ast$.
Consequently, as shown in \Fig{figa2}, the resulting $\TZCC$ are very close but not exactly equal to $\TMPC(n)$.

We now elaborate on the quantitative agreement between $\TMPC(n)$ and $\TMPC(\chi^c)$ apart from the NR regions.
In the NEGF formalism the local charge susceptibility is expressed as
\be \label{def-Tloc-110}
\chi^c
= -\sum_\alpha \frac{\Delta_\alpha}{\Delta + \Delta_p} \int d\omega \, \frac{\partial A(\omega)}{\partial \ep_d} f_{T_{\alpha},\mu_{\alpha}}(\omega),
\ee
and its perturbation by the coupled probe is
\begin{align}
 \delta\chi_p^c &= \chi_p^c(T_p,\mu_p)-\chi_{\rm ref}^c   \nl
 &= - \frac{\Delta_p}{\Delta + \Delta_p} \int d\omega \, \frac{\partial A(\omega)}{\partial \ep_d}
 \big\{f_{T_{p},\mu_{p}}(\omega) \nonumber \\
 &\qquad \qquad -[\zeta_L f_{T_{L},\mu_{L}}(\omega)+\zeta_R f_{T_{R},\mu_{R}}(\omega)] \big\}.   \label{def-Tloc-109}
\end{align}
From \Eqs{zpc-1} and \eqref{def-Tloc-109} it is clear that the probe-induced perturbation to
any local observable $O$ can be cast into a general form of
\be
  \delta O_p = \int d\w \, g_1(\w,T_p)\,g_2^O(\w,\ep_d),  \label{delta-Op-1}
\ee
with
\begin{align}
  g_1(\w,T_p) &= f_{T_{p},\mu_{p}}(\omega)  -[\zeta_L f_{T_{L},\mu_{L}}(\omega)+\zeta_R f_{T_{R},\mu_{R}}(\omega)]  \nl
  &=  \frac{1}{1+e^{\,\omega/T_p}} - \frac{1}{2[1+e^{(\omega+ V/2)/T_L}]}  \nl
  &\quad -\frac{1}{2[1+e^{(\omega- V/2)/T_R}]}  \label{def-Tloc-27}
\end{align}
being a window function determined only by the thermodynamic properties of the leads. Here, the second equality holds because we have $\zeta_L = \zeta_R = \frac{1}{2}$ and $\mu_p = \mu^\ast = 0$ in the case of $\Delta_L = \Delta_R$ and $\mu_R = -\mu_L = \frac{V}{2}$.

In \Eq{delta-Op-1}, the form of the function $g_2^O(\w,\ep_d)$ depends on the definition of the local observable $O$. Specifically, since the spectral function of a noninteracting impurity in the presence of a weakly coupled probe is
\be
  A(\w) = \frac{2}{\pi} \frac{\Delta + \Delta_p}{(\w - \ep_d)^2 + (\Delta + \Delta_p)^2},
\ee
we have
\begin{align}
  g_2^n(\omega,\ep_d) &= \frac{2\Delta_p}{\pi}\frac{1}{(\omega-\epsilon_d)^2+(\Delta+\Delta_p)^2},  \label{g2-n-1} \\
  g_2^{\chi}(\w, \ep_d) &= -\frac{\partial g_2^n}{\partial \ep_d}
  = -\frac{4\Delta_p}{\pi} \frac{\w-\epsilon_d}{[(\omega-\epsilon_d)^2+(\Delta+\Delta_p)^2]^2}.  \label{g2-chi-1}
\end{align}

\begin{figure}[t]
  \centering
  \includegraphics[width=\columnwidth]{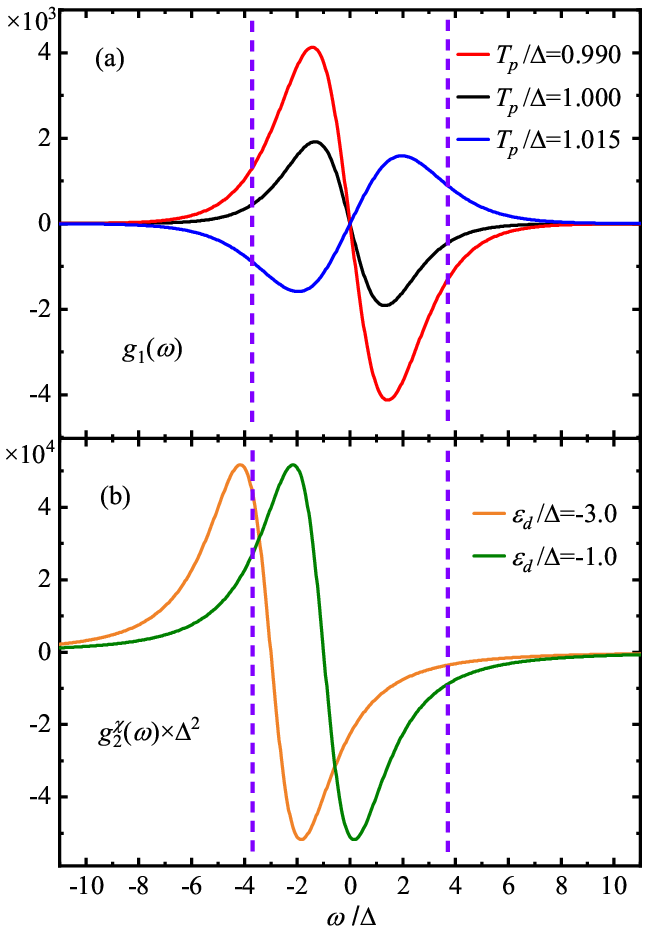}
  \caption {The lineshapes of (a) $g_1(\omega,T_p)$ at various $T_p$ and (b) $g_2^\chi(\omega,\epsilon_d)$ at various $\epsilon_d$ for a noninteracting single-impurity system. The energetic parameters of the system are the same as those adopted in \Fig{figa2}.  The region bounded by the two vertical dashed lines represents an activation energy window, $(\mu_L - \omega_L, \mu_R + \omega_R)$, which is determined by the Fermi distribution function for the leads: $f_{T_\alpha,\mu_\alpha}(\omega) = 1/[e^{(\omega - \mu_\alpha)/T_\alpha} + 1]$.\cite{ye2016thermodynamic} Here, $\omega_\alpha$ is the full width at half maximum of $\frac{\partial}{\partial \omega} f_{T_\alpha,\mu_\alpha}(\omega)$. Under a finite bias voltage, most of the electronic excitations are expected to occur within such an energy window.  } \label{figa3}
  \end{figure}

By combining \Eq{def-Tloc-55} and \Eq{delta-Op-1}, the $\TMPC(O)$ is determined
by tuning $T_p$ until the following ZPC is met:
\be
  \delta O_p =  \int d\w\, g_1(\w, T_p)\, g_2^O(\w, \ep_d) = 0, \label{zpc-2}
\ee
With $T_L = T_R$, $g_1(\w, T_p)$ is an odd function of $\w$, and its significant values appear exclusively within
an activation energy window\cite{ye2016thermodynamic} centered at $\w = 0$; see \Fig{figa3}(a).
%
%

If the ZPC is satisfied, the determined $T^{\ast,{\rm MPC}}(O)$ carries an unambiguous thermodynamic meaning, which is manifested by a correspondence relation,\cite{ye2016thermodynamic} 
i.e., the local observable $O$ of the measured nonequilibrium system is identical
to that of an equilibrium reference system with the temperature being $\TMPC(O)$.

How far the system is from a quantum resonance region can be assessed by analyzing the overlap between the functions $g_1(\omega,T_p)$ and $g_2^\chi(\omega,\epsilon_d)$. For instance, \Fig{figa3}(b) depicts the lineshapes of $g_2^\chi(\omega,\epsilon_d)$ at various $\epsilon_d$ for the single-impurity system studied in \Fig{figa2}. Apparently, when $\epsilon_d$ is close to zero, the function $g_2^\chi(\omega,\epsilon_d)$ falls mainly within the activation window and thus overlaps largely with $g_1(\omega,T_p)$, and the system is considered to be in-resonance. With the $\epsilon_d$ being shifted away from zero, the main part of $g_2^\chi(\omega,\epsilon_d)$ gradually leaves the activation window and thus overlaps less with $g_1(\omega,T_p)$, which implies that the system moves towards the off-resonance region. For the system explored in Figs.~\ref{figa2} and \ref{figa3}, the resonance and off-resonance regions are $|\epsilon_d| < \Delta$ and $|\epsilon_d| > 3\Delta$, and the interstitial regions are the NR regions. A similar analysis can be carried out for a generic quantum impurity system.

{\it Off-resonance--} The impurity system is in the off-resonance region if the impurity energy level $\ep_d$ is
far away from the activation window defined by $g_1(\w, T_p)$.
In such a case, it is the tail of $g_2^O$ that overlaps the main body of $g_1$.
Since $g_2^O(\w, \ep_d)$ varies rather smoothly with $\w$ in the nonequilibrium activation window,
we may use the Taylor expansion and rewrite the ZPC of \Eq{zpc-2} as
\begin{align}
  \delta O_p &= \int d\w \,  g_1(\w, T_p) \, \Big[ g_2^O(0,\ep_d) + \partial_\w g_2^O(0,\ep_d)\, \w  \nl
  & \qquad \qquad + \frac{1}{2}\, \partial_\w^2 g_2^O(\xi,\ep_d)\, \w^2 \Big] \nl
  & = \partial_\w g_2^O(0,\ep_d) \int d\w \, g_1(\w, T_p)\, \w  \nl
  & = 0.     \label{zpc-3}
\end{align}
Here, $\partial_\w \equiv \frac{\partial}{\partial \w}$ and $\partial_\w^2 \equiv \frac{\partial^2}{\partial \w^2}$,
and $\partial_\w^2 g_2^O(\xi,\ep_d)$ with $\xi \in (0,\w)$ is the Lagrange remainder.
The first equality uses the fact that $g_1(\w, T_p)$ is an odd function of $\w$.
It is thus evident that, for the single-impurity system under study, the ZPC holds {\it universally} for any local observable
in the off-resonance region, i.e., $\TMPC(n) = \TMPC(\chi^c) = \TMPC(O)$ for any $O$.

{\it In-resonance--} In contrast, the impurity system is in the resonance region if $\ep_d$ is close to the lead chemical potential
and thus lies within the nonequilibrium activation window. In this case, the value of $\TMPC(O)$ may vary with the specific choice of $O$,
since different local observables may respond differently to nonlocal excitations.
Instead, we still see $\TMPC(n) \approx \TMPC(\chi^c)$ in \Fig{figa2}. This is because of
the following relation resulting from the Taylor expansion and the first equality of \Eq{g2-chi-1},
\begin{align}
  g_2^n(\w, \ep_d) &= g_2^n(\w, 0) - \ep_d \, g_2^\chi(\w, \xi_1) \nl
  &= g_2^n(\w, 0) - \ep_d \, g_2^\chi(\w, \ep_d) \nl
  & \quad - \ep_d (\xi_1 - \ep_d)\, \partial_{\ep_d}g_2^\chi(\w, \xi_2). \label{g2n-w-epd-1}
\end{align}
Here, $\xi_1 \in (0, \ep_d)$ and $\xi_2 \in (\xi_1, \ep_d)$. $g_2^n(\w, 0)$ is an even function of $\w$
and its overlap integral with $g_1(\w, T_p)$ is zero.
The last term on the RHS of \Eq{g2n-w-epd-1} is negligibly small since $\ep_d (\xi_1 - \ep_d) \sim \mathcal{O}(\ep_d^2)$ and $\partial_{\ep_d}g_2^\chi(\w,\xi_2)$ is nearly an even function of $\w$.
Therefore, by combining \Eq{zpc-2} and \Eq{g2n-w-epd-1}, we find that the ZPC for $n$ is approximately equivalent to the ZPC for $\chi^c$,
and hence $\TMPC(n) \approx \TMPC(\chi^c)$.

{\it Near-resonance--} In the NR region, the product of $g_1(\w, T_p)$ and $g_2^O(\w,\ep_d)$ depends sensitively on the nature of $O$, and so is the value of $\TMPC$; see \Fig{figa2}.

From the above theoretical analysis we can conclude that the choice of local observable has little influence on $T^{\ast,{\rm MPC}}$ in the off-resonance regions. In contrast, in the resonance or NR region, the value of $T^{\ast,{\rm MPC}}$ depends on how significantly the local observable is affected by the emerging nonlocal excitations and how sensitively it varies with $T_p$.

When quantum resonances come into play, the $T^{\ast,{\rm MPC}}$ measured by monitoring a suitable local observable, such as the $T^{\ast,{\rm MPC}}(\chi^c)$ depicted in \Fig{figa2}, is {\it still} capable of identifying and quantifying the magnitude of the nonlocal excitations.

\section{Effect of quantum resonances on local temperatures of multi-impurity systems}
\label{thsec-3}

\subsection{Local minimal-perturbation condition}
\label{thsec-3A}

\begin{figure}[t]
\centering
\includegraphics[width=\columnwidth]{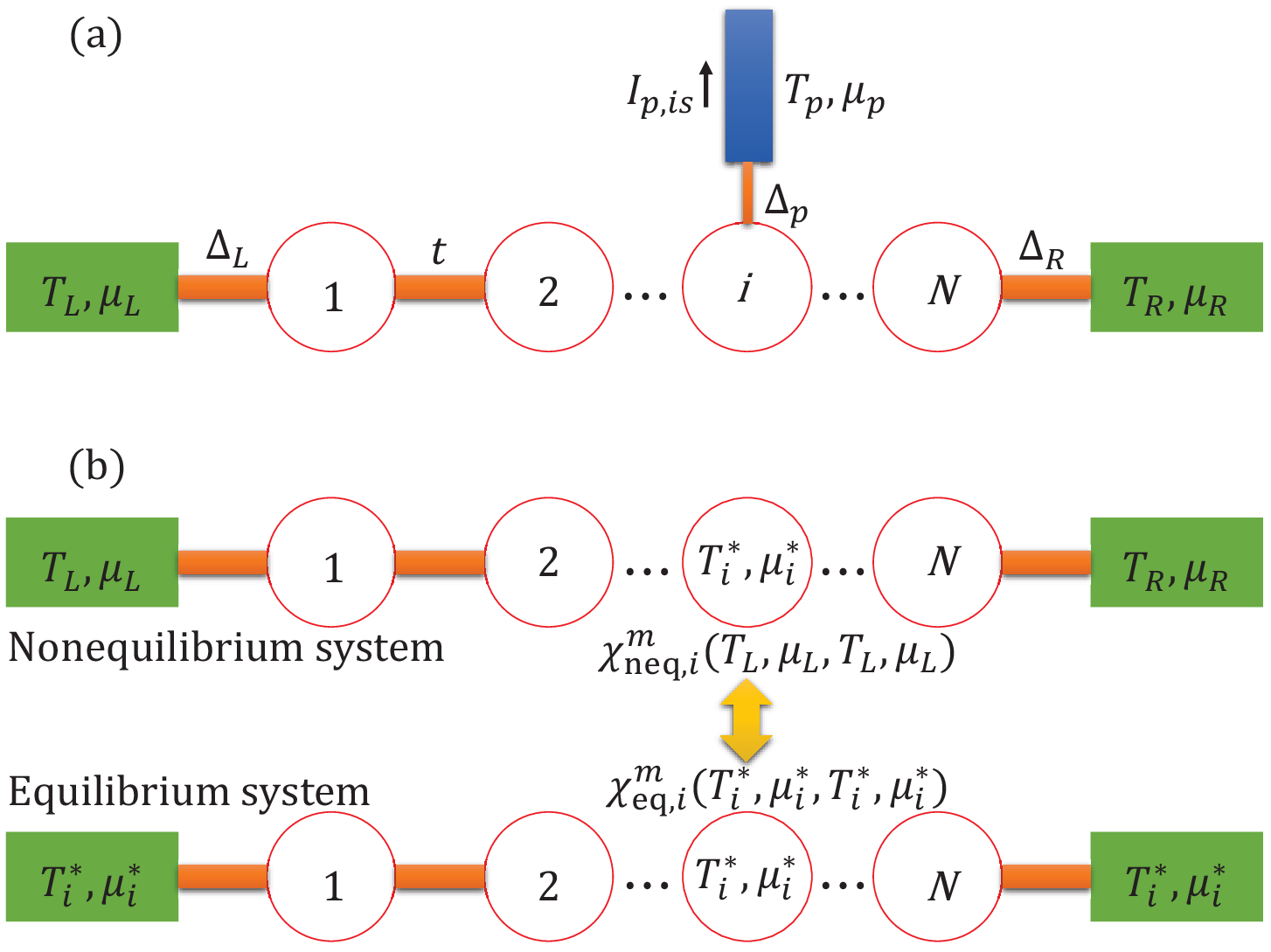}
\caption{(a) Schematic illustration of a serially coupled $N$-impurity system.
The first ($N$th) impurity is coupled to the left (right) lead.
The probe is weakly coupled to the $i$th impurity under study, and the spin-specific
electric current flowing into the probe $I_{p,is}$ is monitored.
(b) Schematic illustration of the correspondence relation which states that
the local magnetic susceptibility of the $i$th impurity in a nonequilibrium system
is equal to that in an equilibrium system, provided that they have the same
local temperature and local chemical potential.
}\label{figa4}
\end{figure}

We now extend the MPC to the systems consisting of more than one impurity.
For numerical convenience, a serially coupled noninteracting $N$-impurity system is considered, which is described by an AIM with
%
\begin{align}
\hat H_{\rm imp} &= \sum_{i=1}^N \epsilon_i \, \hat{n}_i 
 + \sum_{i=1}^{N-1} \big[ t\,(\hat{a}_{i \uparrow}^{\dag} \hat{a}_{{i+1} \uparrow} + \hat{a}_{i \downarrow}^{\dag} \hat{a}_{{i+1} \downarrow}) + {\rm H.c.} \big],  \label{Himp-multi-1}
\end{align}
%
where $\ep_i$ is the on-site energy of the $i$th impurity, and $t$ is the coupling strength between two adjacent impurities.
As illustrated in \Fig{figa4}(a), the $N$-impurity system forms a linear chain, in which the left (right) lead is coupled only to the
first ($N$th) impurity with the coupling strength being $\Delta_{L}$ ($\Delta_{R}$).

In principle the MPC of \Eq{def-Tloc-5} can be formally extended to determine the local temperature and local chemical potential of each individual impurity as follows,
\begin{equation} \label{def-Tloc-66}
\left\{
\begin{aligned}
& {I_{p,i}(T^\ast_i, \mu^\ast_i)= 0}  , \\
&(T^\ast_i, \mu^\ast_i) = {\arg\min_{(T_p, \mu_p)}|\delta O_{p,i}(T_p, \mu_p)|}
\end{aligned}
\right. .
\end{equation}
Here, the probe is weakly coupled to the $i$th impurity. As a natural extension of the MPC, \Eq{def-Tloc-66} is referred to as the local MPC (LMPC).
However, in practice the extension from MPC to LMPC is not always straightforward.
This is because it is often difficult to acquire the minimally perturbed value of a particular local observable $O_{{\rm ref},i}$.
To circumvent this problem, we can choose a local observable whose reference value is known by the intrinsic symmetry
of the system.

For instance, if the investigated multi-impurity system is spin unpolarized, i.e, all the energetic parameters in
\Eq{Himp-multi-1} are spin independent, by coupling a probe to an impurity, the electric current through the probe should also be spin unpolarized.
In other words, if the local observable $O$ is chosen to be the magnetic susceptibility of the electric current through the coupled probe,
$\chi^I_{p,i} \equiv \frac{\partial I_{pz,i}}{\partial H_z}|_{H_z \rightarrow 0}$
with $I_{pz,i} = \frac{1}{2} (I_{p,i\uparrow} - I_{p,i\downarrow})$,
its minimally perturbed value is just $\chi^I_{{\rm ref},i} = 0$, if the $i$th impurity is spin unpolarized in the presence of the probe.
Note that for the measurement of $(T^\ast_i, \mu^\ast_i)$ the magnetic field $H_z$ is applied exclusively on the $i$th impurity.
In contrast, it is much harder to determine the value of $\chi^m_{{\rm ref}, i}$ directly for the $i$th impurity. 

For a single-impurity system, the thermodynamic meaning of $\TMPC$ has been elucidated
via a correspondence condition between the nonequilibrium system under study and a
reference system in thermal equilibrium, i.e., $O_{\rm neq} = O_{\rm eq}$, provided that the $\TMPC(O)$ and $\mu^{\ast,{\rm MPC}}(O)$
of the nonequilibrium system coincide with the equilibrium temperature and chemical potential of the reference system.\cite{ye2016thermodynamic}

In the following, we demonstrate that a correspondence relation can also be established for a multi-impurity system
with the $\chi^I_{p,i}$ chosen as the local observable.

To facilitate the theoretical analysis, we consider a serially coupled noninteracting double-impurity system.
By applying a local magnetic field $H_z$ to the $i$th impurity, the impurity level
is subject to a Zeeman splitting which is assumed to be linearly proportional to $H_z$.
Consequently, for the spin-unpolarized system under study, we have
$\chi^I_{p,i} = \mathcal{C} \frac{\partial I_{p,i \uparrow}}{\partial \ep_i}$
with $\mathcal{C}$ being a constant.
Without loss of generality, the probe is coupled to the first impurity. In the wide-band limit,
the probe-induced perturbation to $\chi^I_{p,1}$ is expressed as
\begin{align}  \label{MPC-dot1}
\delta\chi^I_{p,1} 
&= -\mathcal{C}\, \frac{i}{\pi}  \int d\omega\, \Gamma_p(\omega) \, \partial_{\ep_1} \Big\{ G^<_{\uparrow,11}(\omega) \nl
&  \qquad\qquad + 2i f_{T_p,\mu_p}(\omega) \, {\rm Im}[G^r_{\uparrow, 11}(\omega)] \Big\} \nl
& = \mathcal{C} \,\frac{2 \D_{p}}{\pi} \int d\w\  \Big\{ \D_{L} \left[(\w-\ep_{2})^2+\D_{R}^2\right]  \nl
&\quad \times [f_{T_L,\mu_L}(\w) -f_{T_p,\mu_p}(\w)] + t^2 \D_{R}  \nonumber \\
& \quad  \times  [f_{T_R,\mu_R}(\w) -f_{T_p,\mu_p}(\w)] \Big\} \,  \partial_{\ep_{1}} |B_{p1}(\w)|^2 ,
\end{align}
where
\begin{equation}
B_{p1}(\omega)=\frac{1}{[\omega-\epsilon_{1}+i (\Delta_{L}+\Delta_{p})](\omega-\epsilon_{2}+i \Delta_{R})-t^2}. \label{A-7}
\end{equation}

Similarly, for a spin-unpolarized system, the local magnetic susceptibility of the $i$th impurity can be
rewritten as $\chi^m_{i} = \mathcal{C}' \frac{\partial n_{i \uparrow}} {\partial \ep_i}$,
with $\mathcal{C}'$ being a constant different from $\mathcal{C}$.
In the nonequilibrium steady state characterized by the temperatures and chemical potentials of the left and right leads,
$(T_L,\mu_L,T_R,\mu_R)$, the value of $\chi^m_{1}$ in the absence of the probe is
\begin{align}
\chi^m_{{\rm neq},1} 
&=  \mathcal{C'} \frac{1}{2\pi i}
\int d\omega\,  \, \partial_{\ep_1} [G^<_{\uparrow,11}(\omega)] \nl
&= \mathcal{C'}\, \frac{1}{\pi} \int d\w \, \Big\{ \D_{L} \left[(\w-\ep_{2})^2+\D_{R}^2\right] f_{T_L,\mu_L}(\w) \nonumber \\
& \qquad \qquad + t^2 \D_{R}f_{T_R,\mu_R}(\w)\Big\} \, \partial_{\ep_{1}} | B(\w)|^2,  \label{chi-m1-neq}
\end{align}
where $B(\w) = B_{p1}(\w)|_{\Delta_p = 0}$.
For the reference system in a thermal equilibrium state
characterized by the background temperature $T^\ast_1$ and chemical potential $\mu^\ast_1$,
the corresponding $\chi^m_{1}$ is expressed in a form similar to \Eq{chi-m1-neq},
but with $(T_L,\mu_L,T_R,\mu_R)$ replaced by $(T^\ast_1,\mu^\ast_1,T^\ast_1,\mu^\ast_1)$.
Therefore, the difference between $\chi^m_{{\rm neq},1}$ and $\chi^m_{{\rm eq},1}$ is
\begin{align}
& \chi^m_{{\rm neq},1} (T_L,\mu_L,T_R,\mu_R)  - \chi^m_{{\rm eq},1} (T^\ast_1,\mu^\ast_1,T^\ast_1,\mu^\ast_1) \nonumber  \\
= &\frac{\mathcal{C'}}{\pi} \int d\w\ \Big\{ \D_{L} \left[(\w-\ep_{2})^2+\D_{R}^2 \right]
 \left[ f_{T_L,\mu_L}(\w) - f_{T^\ast_1,\mu^\ast_1}(\w) \right] \nonumber \\
 & \quad +  t^2 \D_{R} \left[f_{T_R,\mu_R}(\w) - f_{T^\ast_1,\mu^\ast_1}(\w) \right] \Big\} \,
   \partial_{\ep_{1}}| B(\w)|^2.  \label{deviation-dot1}
\end{align}
By comparing \Eq{MPC-dot1} and \Eq{deviation-dot1}, it is easy to recognize that the relation
\be
 \chi^m_{{\rm neq},1}(T_L,\mu_L,T_R,\mu_R) = \chi^m_{{\rm eq},1} (T_1^\ast,\mu_1^\ast,T_1^\ast,\mu_1^\ast)  \label{corrspondence-dot1}
\ee
holds provided that
\be
  \left. \frac{\delta\chi^I_{p,1} (T_1^\ast,\mu_1^\ast)}{\D_{p}} \right|_{\D_{p} \to 0} = 0.
  \label{lmpc-1}
\ee
A similar relation can be established for the second impurity of the double-impurity system,
or any impurity of a generic multi-impurity system; see \Fig{figa4}(b).

Equation~\eqref{lmpc-1} is the local ZPC for the local observable $\chi^I_{p,1}$,
and the thermodynamic meaning of the resulting ($T^\ast_1, \mu^\ast_1$)
is unambiguously given by \Eq{corrspondence-dot1}.
When the local ZPC of \Eq{lmpc-1} cannot be reached, such as in the NR region,
the LMPC of \Eq{def-Tloc-66} with $O_i = \chi^I_{p,i}$ still yields a unique $T^\ast_i$
which could characterize the emergence of quantum resonance effects;
see \Sec{thsec-3B} and \Sec{subsec:chain} for details.

\subsection{Validity of LMPC for single impurity systems} \label{thsec-3B}

\begin{figure}[t]
\centering
\includegraphics[width=\columnwidth]{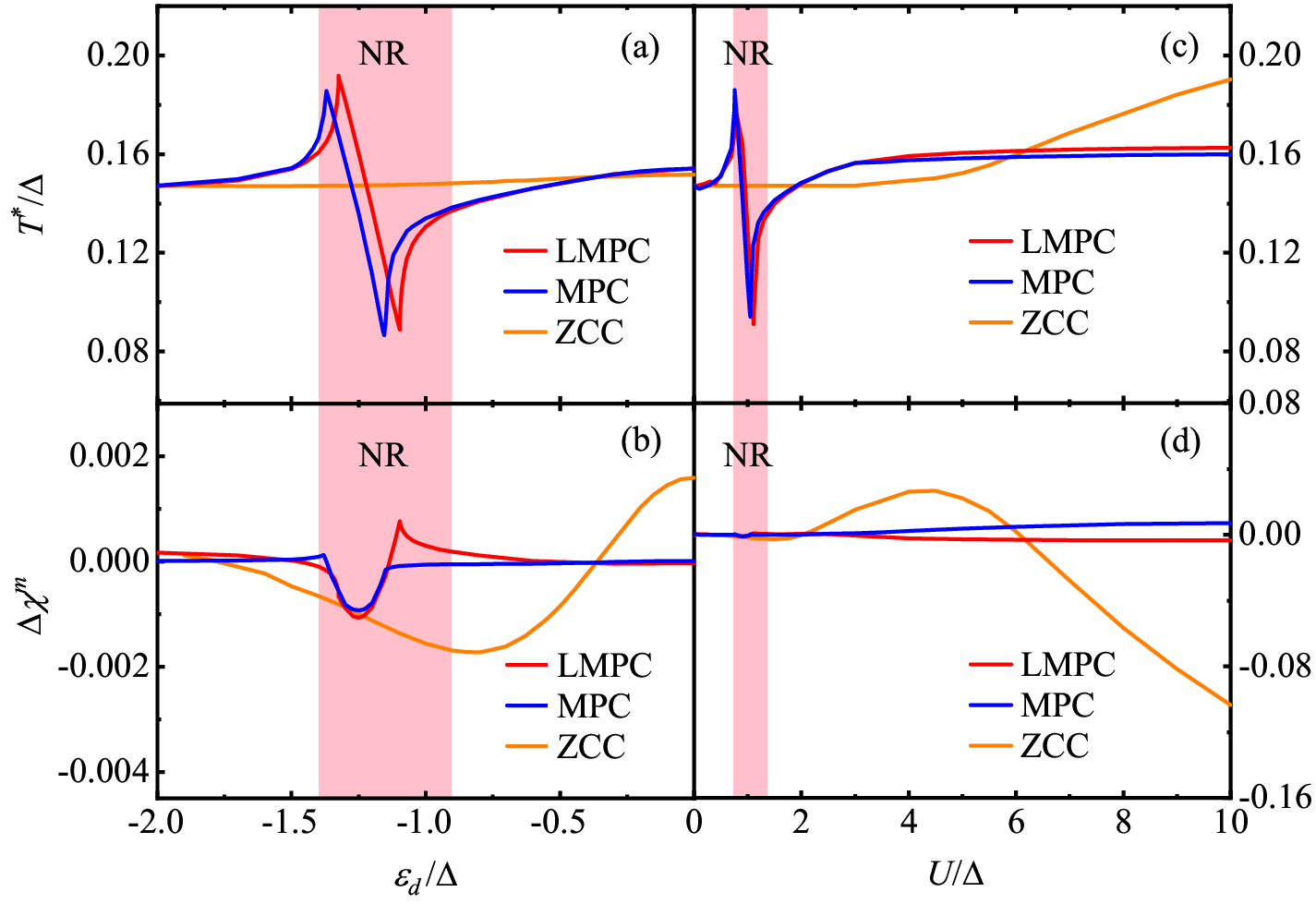}
\caption{(a) $T^\ast$ determined by the ZCC, MPC (with $O = \chi^m$), and LMPC (with $O = \chi^I_p$)
as functions of $\ep_d$ for a noninteracting single-impurity system under an antisymmetric bias voltage of $\mu_R = -\mu_L = V/2 =0.2\Delta$.
(b) The relative deviation from the correspondence relation, $\Delta\chi^m = \chi^m_{\rm eq} /\chi^m_{\rm neq} -1$, versus $\ep_d$.
(c) $T^\ast$ determined by the ZCC, MPC, and LMPC as functions of $U$ for an interacting single-impurity system with $\epsilon_d=-2\Delta$
under the same antisymmetric bias voltage.
(d) The relative deviation $\Delta \chi^m$ versus $U$.
Other energetic parameters adopted are (in units of $\Delta$): $T_L = T_R = 0.1$, $\Delta_L =\Delta_R = 0.5$, and $W_L = W_R = 20$.
The shaded areas represent the NR regions.} \label{figa5}
\end{figure}

Before applying the LMPC-based protocol to multi-impurity systems, we first
examine its consistency with the MPC-based protocol
for single-impurity systems. In principle, the LMPC-based protocol with $\chi^I_p$ as the local observable
is equivalent to the MPC-based definition with $O = \chi^m$.
This is because they both lead to the correspondence relation
$\chi^m_{{\rm neq}}(T_L,\mu_L,T_R,\mu_R) = \chi^m_{{\rm eq}} (T^\ast,\mu^\ast,T^\ast,\mu^\ast)$,
provided that the ZPC for the local observable can be achieved.
Nevertheless, for a real system the equality in the above relation is subject to a small error due to the finite band width of the leads, which could affect the two local observables somewhat differently. Consequently, in \Fig{figa5} the resulting $\TMPC$ and $T^{\ast, {\rm LMPC}}$ exhibit some minor deviation.

\Figure{figa5}(a) shows the $T^\ast$ of a noninteracting single-impurity system under an antisymmetric bias voltage
as a function of $\epsilon_d$.
It is found that, while the $T^{\ast,{\rm LMPC}}$ agree closely with the $T^{\ast,{\rm MPC}}$ outside the NR region,
they display an appreciable difference in the NR region despite the overall similar lineshape.

In the NR region, if the value of the monitored local observable (such as $\chi^I_p$ and $\chi^m$) is
strongly affected by the emergence of quantum resonances,
it could be difficult to reach the ZPC by simply tuning the $T_p$.
In such a case, $T^\ast$ has to be determined by searching for the $T_p$ that yields a minimal nonzero perturbation to $O$ ($\delta O_p$).
Thus, the resulting $T^{\ast,{\rm LMPC}}$ or $T^{\ast,{\rm MPC}}$ may exhibit large oscillations and
deviate from each other
because the minimal $\delta O_p$ tends to vary sensitively with the nonlocal excitations introduced by the emerging quantum resonance.

Figure~\ref{figa5}(b) depicts the relative deviation of $\chi^m$ of the nonequilibrium impurity system
from that of the reference equilibrium system, $\Delta \chi^m = \chi^m_{\rm eq} / \chi^m_{\rm neq} - 1$.
Evidently, while $\Delta \chi^m$ almost vanishes with either $T^{\ast,{\rm MPC}}$ or $T^{\ast,{\rm LMPC}}$,
it remains of a finite magnitude in the NR region where the ZPC cannot be reached. 
In contrast, the $\TZCC$ are almost constant in the whole range of $\ep_d$ with a considerably larger $\Delta \chi^m$,
and they show no sign of quantum resonances at all.

Figures~\ref{figa5}(c) and~\ref{figa5}(d) depict the $T^\ast$ and $\Delta \chi^m$ of an interacting single-impurity system
under an antisymmetric bias voltage as a function of $U$, respectively.
Similar to the case of a noninteracting impurity, as long as the ZPC can be satisfied,
the $T^{\ast,{\rm LMPC}}$ and $T^{\ast,{\rm MPC}}$ agree closely with each other, otherwise they display a minor difference.
%
It is worth pointing out that the low background temperature 
enables the formation of Kondo states,\cite{Yigal1993Low} 
which provide resonant channels for electrons to transport across the impurity.
Therefore, the system remains in the resonance region with a sufficiently large $U$ ($U > -\ep_d$).
Again, the $\TZCC$ vary rather smoothly and do not reflect the formation of quantum resonant states at all.

The above results clearly verify that the newly proposed LMPC is generally consistent with the original MPC for single-impurity systems.

\subsection{Local temperatures of multi-impurity systems
and the effect of quantum resonances} \label{subsec:chain}

\begin{figure}[t]
\centering
\includegraphics[width=\columnwidth]{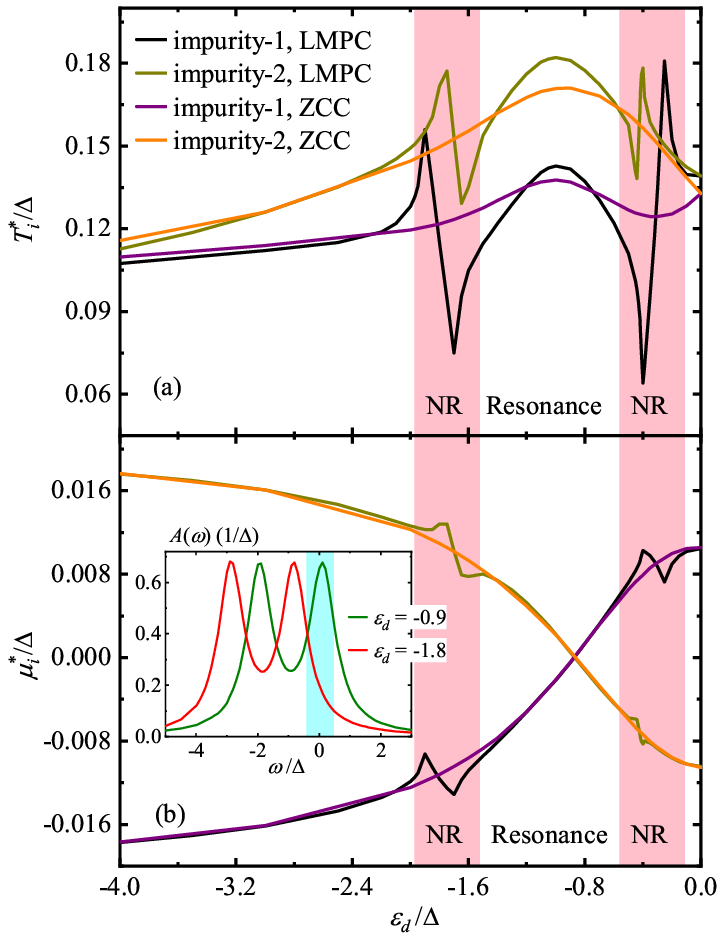}
\caption{Evolution of (a) $T^\ast_i$ and (b) $\mu^\ast_i$ determined by the ZCC and the LMPC with the variation of $\ep_i = \ep_d$
for a noninteracting double-impurity system under an antisymmetric bias voltage of $\mu_R=-\mu_L=V/2 = 0.2\Delta$.
Other energetic parameters adopted are (in units of $\Delta$): $T_L=T_R=0.1$, $U = 0$, $t=1$, $\Delta_{L}=\Delta_{R}=0.5$, and $W_L=W_R=20$.
The shaded areas in the main panels represent the NR regions.
The inset of (b) depicts the total spectral function of the two impurities $A(\w)$ at different $\ep_d$,
where the shaded area indicates the nonequilibrium activation window.
}\label{figa6}
\end{figure}

We now employ the LMPC-based protocol to investigate the distribution of local temperatures in
a double-impurity system under an antisymmetric bias voltage.
Here, the two impurities are presumed to have the same onsite energy, i.e., $\ep_i = \ep_d$.

\Figure{figa6} depicts the evolution of $(T^\ast_i, \mu^\ast_i)$ of the two impurities
with the variation of $\ep_d$.
In analogy with the case of single-impurity systems, while the $\TiLMPC$ agree well with the $\TZCC_i$ in the absence of resonance,
they are distinctly different in the two NR regions.

It is worth noting that $T^\ast_1 < T^\ast_2$ at almost any $\ep_d < 0$, which can be explained as follows.
With $\ep_d < 0$ the total spectral function, $A(\w)$, of the two impurities has a distribution more on the negative energy side,
and this means that the double-impurity system has a positive Seebeck coefficient $S$.\cite{Don0211747,Ye2014Thermopower}
Consequently, the voltage-generated heat current between the two impurities follows the opposite direction of the electric current, i.e., from left to right.
Such a heat current thus creates an internal thermal gradient across the two impurities with $T^\ast_1 < T^\ast_2$.

At $\ep_d = 0$, $A(\w)$ becomes an even function of $\w$.
As a result we have $S=0$, and hence the voltage-generated internal thermal gradient also becomes zero, i.e., $T^\ast_1 = T^\ast_2$.
This is indeed confirmed by our calculation results shown in \Fig{figa6}(a).
Furthermore, it is also inferred that $T^\ast_1 > T^\ast_2$ at $\ep_d > 0$ (data not shown). 

It is also interesting to observe that, while the left (right) lead has a lower (higher)
chemical potential, the $\mu^\ast_i$ of the neighboring impurity is not necessarily lower (higher); see \Fig{figa6}(b).
The fluctuation of $\mu^\ast_i$ manifests the quantum coherence nature of the electron transport driven by the bias voltage.
In particular, $\mu^\ast_1 = \mu^\ast_2 = 0$ at $\ep_d = -0.9\Delta$,
where a resonant state resides right at the center of the nonequilibrium activation window; see the $A(\w)$ in the inset of \Fig{figa6}(b).
The uniformity of $\mu^\ast_i$ indicates that the voltage-driven excitations are predominantly nonlocal
as they occur via the resonant state which involves both impurities.

In \Fig{figa6}, it is again apparent that the $(T^\ast_i, \mu^\ast_i)$ predicted by the ZCC vary smoothly with $\ep_d$,
and completely neglect the existence of nonlocal excitations;
whereas those determined by the LMPC exhibit large oscillations in the NR regions, which clearly accentuates the emergence of a quantum resonance.

\begin{figure}[t]
\centering
\includegraphics[width=\columnwidth]{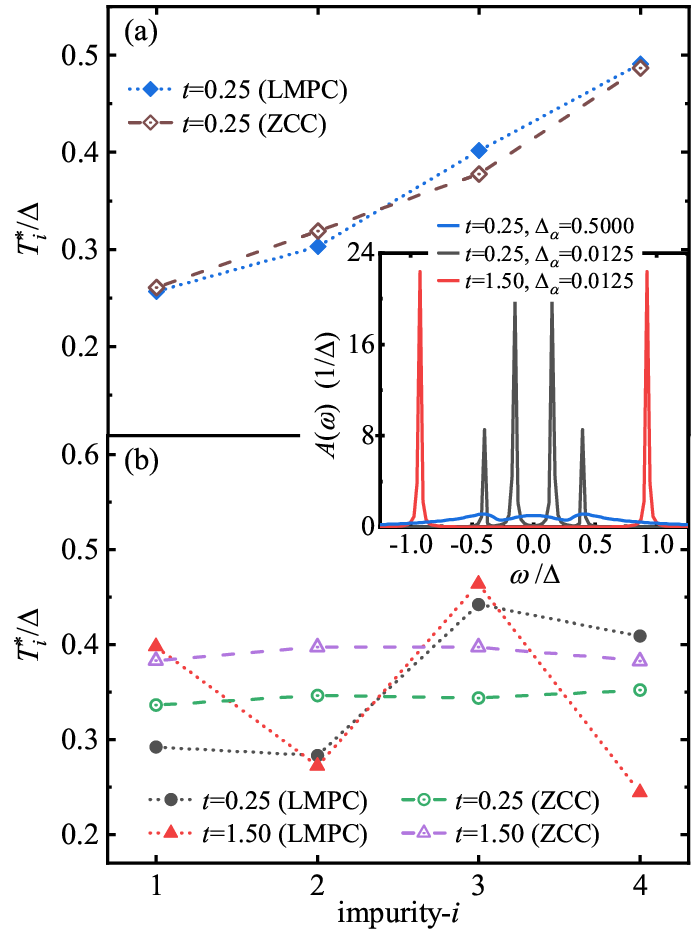}
\caption{Local temperature profile $T^\ast_i$ determined by the ZCC and the LMPC for a noninteracting four-impurity chain
under a thermal bias with (a) a strong impurity-lead coupling of $\Delta_\alpha = 0.5\Delta$ ($\alpha = L,R$)  and
(b) a weak impurity-lead coupling of $\Delta_\alpha = 0.0125\Delta$.
Other energetic parameters adopted are (in units of $\Delta$): $\epsilon_i = \ep_d = 0$, $U = 0$, $\mu_L = \mu_R = 0$,
$T_L = 0.25$, $T_R = 0.5$, and $W_L = W_R = 5$.
The inset depicts the system spectral function $A(\w)$ at different $t$ and $\Delta_\alpha$.
}\label{figa7}
\end{figure}

We proceed to study a linear chain comprised of four noninteracting impurities subject to a thermal bias, i.e., $T_L < T_R$.
The local chemical potential on each impurity $\mu^\ast_i$ is nearly zero due to the absence of bias voltage.
Figure~\ref{figa7} depicts the distribution of $T^\ast_i$ along the chain determined by the ZCC and the LMPC
for various values of $t$ and $\Delta_\alpha$ ($\alpha=L,R$).

As shown in \Fig{figa7}(a), when the terminal impurities are coupled strongly to the leads,
both the ZCC and the LMPC predict the $T^\ast_i$ vary almost linearly with $i$, i.e.,
the distribution of local temperature along the chain obeys the classical Fourier's law.\cite{Zhang2019local}
Note that here the restoration of the Fourier's law is not because of disorder\cite{Dub09115415} or dephasing caused by an external source,\cite{Dub09042101} which are absent from the AIM under study.
Instead, the linear profile of $T^\ast_i$ is associated with the substantial broadening of the spectral peaks in $A(\w)$.\cite{Inu184304}
This means that the thermal transport process involves electronic states in a wide range of energies.
The phases of these states are averaged out when $T^\ast_i$ are measured, which leads to a classical-like behavior.
As the inset of \Fig{figa7} shows, the impurity-lead coupling $\Delta_\alpha$ ($\alpha= L$ or $R$) affects significantly the sharpness of the peaks, while the coupling strength $t$ between two adjacent impurities has important influence on the distance of neighboring peaks in a system.

In contrast, \Fig{figa7}(b) concerns another scenario in which the impurity-lead coupling is extremely weak,
so that the thermal transport occurs almost exclusively via the quantum resonant states formed on the chain.
In such a scenario, the ZCC and the LMPC yield very different predictions on the distribution of $T^\ast_i$.
Specifically, the $\TZCC_i$ of all four impurities is close to a certain value between $T_L$ and $T_R$,
while the $\TiLMPC$ exhibit large oscillations along the chain, which clearly violates the Fourier's law.

Inui \emph{et al.}~\cite{Inu184304} have reported strong oscillations of temperature distribution
in a graphene flake weakly coupled to the electrodes under a thermal bias due to quantum interference.
But in their study the local temperatures $T^{\ast,{\rm ZCC}}_i$ still remain constant on a
relatively small scale in the weak-coupling regime, similar to the curve of $T^{\ast,{\rm ZCC}}_i$ in  \Fig{figa7}(b).
Note that the chain in our work is very short, so that even with weak impurity-lead couplings
the ZCC-defined $T^\ast_i$ cannot reveal prominent oscillations.
In contrast, the LMPC-defined $T^{\ast,{\rm LMCP}}_i$ oscillations in our work are much more significant.
In the strong-coupling regime, the temperature profile for the graphene flake is much closer to that predicted by classical Fourier's law,\cite{Inu184304} which is consistent with our results in
\Fig{figa7}(a).

\section{Concluding remarks} \label{thconc}

In conclusion, our study has demonstrated that, while the MPC could yield a unique local temperature for a given local observable $\hat{O}$, the value of $\TMPC$ may vary with the specific choice of $\hat{O}$, particularly in the NR region where both the local and nonlocal excitations could take place inside a quantum impurity system.
It is also noticed that, while $\TMPC$ agrees quantitatively to $\TZCC$ away from the in-resonance region,
their difference is appreciable in the NR region. 
This indicates that, when quantum resonant states emerge in the system, it is difficult to fully eliminate the 
influence of the probe on the local observable $O$ by tuning $T_p$ and $\mu_p$, because of the presence of nonlocal excitations.

We have also proposed an operational protocol based on the LMPC, which extends the operational principle of MPC to multi-impurity systems.
%
%
Using the LMPC, we studied the effect of quantum resonances on the
local temperatures of noninteracting multi-impurity systems.
We found that the $T^{\ast,{\rm LMPC}}_i$ of double-impurity systems under an antisymmetric bias voltage agree well with the $T^{\ast,{\rm ZCC}}_i$
in the absence of resonances. On the other hand, they are distinctly different in the two NR regions, which is analogous to the case of single-impurity systems.
Applying the LMPC to a linear chain of four impurities,
we found that the strong quantum resonance effects lead to prominent oscillations in the local temperature,
which are not observed with the ZCC-based definition. 
%


%
It is important to point out that the practical implementation of the MPC- and LMPC-based protocols is very straightforward, as they do not require the direct measurement of heat currents. Moreover, in the existing experimental and theoretical works, the measured local temperature is often associated with concrete physical properties, such as thermoelectricity,\cite{mills1998scanning,luo1996nanofabrication} electrical resistance,\cite{zhang2011batch,pylkki1994scanning} thermal expansion,\cite{nakabeppu1995scanning} fluorescence,\cite{saidi2009scanning} energy reactance,\cite{ludovico2014dynamical,ludovico2018probing} etc. 
In the present work, the local temperature determined by the MPC- or LMPC-based protocol is closely related to the monitored local observable. The measured local temperature can be unambiguously interpreted by the correspondence relation, i.e., the monitored local observable of the nonequilibrium system under study is identical to that of an equilibrium reference system, provided that the perturbation induced by the probe can be completely suppressed by tuning $T_p$ and $\mu_p$. 

\acknowledgments

The authors acknowledge the support from the Ministry of Science and Technology of China (Grant Nos. 2016YFA0400900, 2016YFA0200600, and 2017YFA0204904), the National Natural Science Foundation of China (Grant Nos. 21973086 and 21633006),
and the Ministry of Education of China (111 Project Grant No. B18051).
The computational resources are provided by the Supercomputing Center of University of Science and Technology of China.

\appendix

\section{MPC zero perturbation implies zero energy and particle flows: A formal proof} \label{app:proof}


Consider a system interacting with several environments $\alpha$. The total Hamiltonian is $\HT = \Hs + \sum_\alpha (\hH_\alpha + \hH_{{\rm s} \alpha})$,
where $\Hs$ represents the primary system of interest, $\hH_\alpha$ represents the $\alpha$th environment ($\alpha=p$ labels the probe), and $\hH_{{\rm s} \alpha}$ is the system-environment interaction. 
The density matrix of the total system $\rhot$ satisfies the equation
\be \label{eqn:eom-1}
\dot{\rho}_{_{\rm T}} = -i[\HT, \rhot].
\ee
We define $\rho \equiv \trb (\rhot)$ as the reduced density matrix of the primary system, with $\trb$ denoting a trace over all the environment degrees of freedom. A quantum master equation (QME) for $\rho$ can be formally written as 
\be  \label{eqn:qme-1}
\dot{\rho} = -i[\Hs,\rho] + \sum_\alpha \mathcal{R}_\alpha[\rho],
\ee
where $\mathcal{R}_\alpha[\rho]$ represents the dissipation term between the system and the $\alpha$th environment.

In the following, we give a formal proof showing that 
$\mathcal{R}_\alpha[\rho] = 0$ implies $I_\alpha = 0$ and $J^E_\alpha =0$ for the $\alpha$th reservoir environment 
coupled to an open system at a stationary state, where $I_\alpha$ and $J^E_\alpha$ are the electrical and energy currents 
between the system and the $\alpha$th reservoir, respectively.

By referring to \Eq{eqn:eom-1}, the dissipation term $\mathcal{R}_\alpha[\rho]$ in \Eq{eqn:qme-1} originates from the following expression 
\be
    \mathcal{R}_\alpha[\rho] = -i \, \trb\{ [\hH_{{\rm s} \alpha} + \hH_\alpha, \rhot] \}. 
\ee

Consider first the electric current between the system and the $\alpha$th reservoir, which is defined by
\be  \label{eqn:I-1}
  I_\alpha \equiv \dot{N}_\alpha(t) = \trt \{\hat{N}_\alpha \dot{\rho}_{_{\rm T}} \} 
  = -i \,\la [\hat{N}_\alpha, H_{{\rm s} \alpha}] \ra. 
\ee
Here, $\la \cdots \ra \equiv \trt \{ \cdots \rhot  \}$, and we have used the fact that the number of particles operator $\hat{N}_\alpha$ commutes with $\Hs$ and $\hH_{\alpha'}$. 
On the other hand, the QME for $\rho$ gives rise to a formula for the conservation of particles, 
$\dot{N}_{_{\rm S}}(t) = \trs\{ \hat{N}_{_{\rm S}} \dot{\rho} \} 
= -\sum_\alpha \tilde{I}_\alpha$, where  
\be \label{eqn:tI-1}
  \tilde{I}_\alpha \equiv - \trs\{ \hat{N}_{_{\rm S}} \mathcal{R}_\alpha[\rho]\} 
  = i \,\la  [\hat{N}_{_{\rm S}}, H_{{\rm s} \alpha}] \ra. 
\ee
The last equality of \Eq{eqn:tI-1} uses the fact that $\hat{N}_{_{\rm S}}$ commutes with $\hH_\alpha$. 
By comparing \Eq{eqn:I-1} and \Eq{eqn:tI-1}, we have
\be
 I_\alpha = \tilde{I}_\alpha -i \,\la [\hat{N}_\alpha+\hat{N}_{_{\rm S}}, H_{{\rm s} \alpha}] \ra.
\ee
If every term in the interaction Hamiltonian $\hH_{{\rm s} \alpha}$ conserves the number of particles within the system and the $\alpha$th environment (e.g., in the Anderson impurity model $\hH_{{\rm s} \alpha}$ causes electron transfer only between the impurity and the $\alpha$th reservoir), we have $[\hat{N}_\alpha + \hat{N}_{_{\rm S}}, \hH_{{\rm s} \alpha}] =0$. Thus,  $\mathcal{R}_\alpha[\rho] = 0$ immediately leads to $I_\alpha = \tilde{I}_\alpha = 0$.

Consider then the energy flow between the system and the $\alpha$th environment, which is defined by 
\be  \label{eqn:J-1}
  J^E_\alpha \equiv \dot{E}_\alpha(t) 
  = -i \,\la [\hH_\alpha, \hH_{{\rm s} \alpha}]\ra.
\ee
Similarly, the QME for $\rho$ gives rise to a formula for the conservation of energy, 
$\dot{E}_{_{\rm S}}(t) = \trs\{ \Hs \dot{\rho} \} 
= -\sum_\alpha \tilde{J}^E_\alpha$, where
%
%
\be  \label{eqn:tJ-1}
\tilde{J}^E_\alpha \equiv - \trs\{ \Hs \mathcal{R}_\alpha[\rho]\} = i\, \la [\Hs, \hH_{{\rm s} \alpha}]\ra. 
\ee
Here, we have used the fact that $\Hs$ commutes with $\hH_\alpha$. By comparing \Eq{eqn:J-1} with \Eq{eqn:tJ-1}, we have
\begin{align}  \label{eqn:J-2}
  J^E_\alpha &= \tilde{J}^E_\alpha -i \, \la [\HT, \hH_{{\rm s} \alpha}]\ra - i \sum_{\alpha'\neq \alpha} 
  \trt \{ [\hH_{{\rm s} \alpha}, \hH_{{\rm s} \alpha'}]\, \rhot \}   \nl
  &= - \trs\{ \Hs \mathcal{R}_\alpha[\rho]\}   - \trt \{\hH_{{\rm s} \alpha} \dot{\rho}_{_{\rm T}} \}
  - i \sum_{\alpha'\neq \alpha}  \la [\hH_{{\rm s} \alpha},H_{{\rm s} \alpha'}]\ra. 
\end{align}
Note that $\sum_\alpha \sum_{\alpha'\neq \alpha}  [\hH_{{\rm s} \alpha}, \hH_{{\rm s} \alpha'}] = 0$.

Clearly, $\mathcal{R}_\alpha[\rho] = 0$ alone does not guarantee $J^E_\alpha = 0$. 
Further consideration is needed for the two other terms on the right-hand side of \Eq{eqn:J-2}. 
First, we have $[\hH_{{\rm s} \alpha}, \hH_{{\rm s} \alpha'}] = 0$ if $\hH_{{\rm s} \alpha}$ and $\hH_{{\rm s} \alpha'}$ involve the system's different degrees of freedom; otherwise, $\la \hH_{{\rm s} \alpha} \hH_{{\rm s} \alpha'}\ra$ represents the covariance between a stochastic variable of the 
$\alpha$th environment and a stochastic variable of the $\alpha'$th environment. Such a covariance is usually zero because the two environments are statistically independent. 
Moreover, the term $\trt \{\hH_{{\rm s} \alpha} \dot{\rho}_{_{\rm T}} \}$ can be interpreted as the rate at which the interaction energy between the system and the $\alpha$th environment [$E_{{\rm s} \alpha}(t)$] varies with time. Such a rate is zero when the total system reaches a stationary state, (cf. Fig.~1 of Ref.~\onlinecite{Son17064308}). 
We thus conclude that $\mathcal{R}_\alpha[\rho] = 0$ leads to  $J^E_\alpha = 0$ for the system at a stationary state. 

Finally, the heat current between the system and the $\alpha$th reservoir is also zero, i.e., $J^H_\alpha = J^E_\alpha - \mu_\alpha I_\alpha = 0$.

%
\bibliographystyle{aip}
\bibliography{bibrefs}

\begin{thebibliography}{10}

\bibitem{Zhang2019local}
D.~Zhang, X.~Zheng, and M.~{Di Ventra},
\newblock Phys. Rep. {\bf 830}, 1 (2019).

\bibitem{Hof09779}
E.~A. Hoffmann, H.~A. Nilsson, J.~E. Matthews, N.~Nakpathomkun, A.~I. Persson,
  L.~Samuelson, and H.~Linke,
\newblock Nano Lett. {\bf 9}, 779 (2009).

\bibitem{menges2013thermal}
F.~Menges, H.~Riel, A.~Stemmer, C.~Dimitrakopoulos, and B.~Gotsmann,
\newblock Phys. Rev. Lett. {\bf 111}, 205901 (2013).

\bibitem{Men1610874}
F.~Menges, P.~Mensch, H.~Schmid, H.~Riel, A.~Stemmer, and B.~Gotsmann,
\newblock Nat. Commun. {\bf 7}, 10874 (2016).

\bibitem{Inu184304}
S.~Inui, C.~A. Stafford, and J.~P. Bergfield,
\newblock ACS Nano {\bf 12}, 4304 (2018).

\bibitem{Lee13209}
W.~Lee, K.~Kim, W.~Jeong, L.~A. Zotti, F.~Pauly, J.~C. Cuevas, and P.~Reddy,
\newblock Nature (London) {\bf 498}, 209 (2013).

\bibitem{Cui18122}
L.~Cui, R.~Miao, K.~Wang, D.~Thompson, L.~A. Zotti, J.~C. Cuevas, E.~Meyhofer,
  and P.~Reddy,
\newblock Nat. Nanotechnol. {\bf 13}, 122 (2018).

\bibitem{Nov19016806}
D.~Novko, J.~C. Tremblay, M.~Alducin, and J.~I. Juaristi,
\newblock Phys. Rev. Lett. {\bf 122}, 016806 (2019).

\bibitem{Zei04871}
M.~P. Zeidler, C.~Tan, Y.~Bellaiche, S.~Cherry, S.~H\"{a}der, U.~Gayko, and
  N.~Perrimon,
\newblock Nat. Biotechnol. {\bf 22}, 871 (2004).

\bibitem{Kuc1354}
G.~Kucsko, P.~C. Maurer, N.~Y. Yao, M.~Kubo, H.~J. Noh, P.~K. Lo, H.~Park, and
  M.~D. Lukin,
\newblock Nature (London) {\bf 500}, 54 (2013).

\bibitem{He1626737}
Y.-M. He and B.-G. Ma,
\newblock Sci. Rep. {\bf 6}, 26737 (2016).

\bibitem{Sadat2012High}
S.~Sadat, E.~Meyhofer, and P.~Reddy,
\newblock Rev. Sci. Instrum. {\bf 83}, 084902 (2012).

\bibitem{mecklenburg2015Nanoscale}
M.~Mecklenburg, W.~A. Hubbard, E.~R. White, R.~Dhall, S.~B. Cronin, S.~Aloni,
  and B.~C. Regan,
\newblock Science {\bf 347}, 629 (2015).

\bibitem{Gro11287}
K.~L. Grosse, M.-H. Bae, F.~Lian, E.~Pop, and W.~P. King,
\newblock Nat. Nanotechnol. {\bf 6}, 287 (2011).

\bibitem{Gurrum2005Scanning}
S.~P. Gurrum, Y.~K. Joshi, W.~P. King, and K.~Ramakrishna,
\newblock J. Heat Transfer. {\bf 127}, 809 (2005).

\bibitem{di2008electrical}
M.~Di~Ventra,
\newblock {\em Electrical Transport in Nanoscale Systems},
\newblock Cambridge University Press, Cambridge, 2008.

\bibitem{Liu2015Density}
S.~Liu, A.~Nurbawono, and C.~Zhang,
\newblock Sci. Rep. {\bf 5}, 15386 (2015).

\bibitem{Hua061240}
Z.~Huang, B.~Xu, Y.~Chen, M.~{Di Ventra}, and N.~Tao,
\newblock Nano Lett. {\bf 6}, 1240 (2006).

\bibitem{huang2007local}
Z.~Huang, F.~Chen, R.~D'agosta, P.~A. Bennett, M.~Di~Ventra, and N.~Tao,
\newblock Nat. Nanotechnol. {\bf 2}, 698 (2007).

\bibitem{Kim14203107}
K.~Kim, W.~Jeong, W.~Lee, S.~Sadat, D.~Thompson, E.~Meyhofer, and P.~Reddy,
\newblock Appl. Phys. Lett. {\bf 105}, 203107 (2014).

\bibitem{Thi15854}
H.~Thierschmann, R.~S{\'a}nchez, B.~Sothmann, F.~Arnold, C.~Heyn, W.~Hansen,
  H.~Buhmann, and L.~W. Molenkamp,
\newblock Nat. Nanotechnol. {\bf 10}, 854 (2015).

\bibitem{Ivo16014301}
I.~Lon\v{c}ari\'{c}, M.~Alducin, P.~Saalfrank, and J.~I. Juaristi,
\newblock Phys. Rev. B {\bf 93}, 014301 (2016).

\bibitem{Idr18095901}
J.~C. Idrobo, A.~R. Lupini, T.~Feng, R.~R. Unocic, F.~S. Walden, D.~S.
  Gardiner, T.~C. Lovejoy, N.~Dellby, S.~T. Pantelides, and O.~L. Krivanek,
\newblock Phys. Rev. Lett. {\bf 120}, 095901 (2018).

\bibitem{scovil1959three}
H.~E.~D. Scovil and E.~O. Schulz-DuBois,
\newblock Phys. Rev. Lett. {\bf 2}, 262 (1959).

\bibitem{curzon1975efficiency}
F.~L. Curzon and B.~Ahlborn,
\newblock Am. J. Phys. {\bf 43}, 22 (1975).

\bibitem{lieb1999physics}
E.~H. Lieb and J.~Yngvason,
\newblock Phys. Rep. {\bf 310}, 1 (1999).

\bibitem{allahverdyan2001breakdown}
A.~E. Allahverdyan and T.~M. Nieuwenhuizen,
\newblock Phys. Rev. E {\bf 64}, 056117 (2001).

\bibitem{casas2003temperature}
J.~Casas-V{\'a}zquez and D.~Jou,
\newblock Rep. Prog. Phys. {\bf 66}, 1937 (2003).

\bibitem{kieu2004second}
T.~D. Kieu,
\newblock Phys. Rev. Lett. {\bf 93}, 140403 (2004).

\bibitem{bustamante2005nonequilibrium}
C.~Bustamante, J.~Liphardt, and F.~Ritort,
\newblock Phys. Today {\bf 58}, 43 (2005).

\bibitem{horhammer2008information}
C.~H{\"o}rhammer and H.~B{\"u}ttner,
\newblock J. Stat. Phys. {\bf 133}, 1161 (2008).

\bibitem{bergfield2009many}
J.~P. Bergfield and C.~A. Stafford,
\newblock Phys. Rev. B {\bf 79}, 245125 (2009).

\bibitem{levy2012quantum}
A.~Levy, R.~Alicki, and R.~Kosloff,
\newblock Phys. Rev. E {\bf 85}, 061126 (2012).

\bibitem{horodecki2013fundamental}
M.~Horodecki and J.~Oppenheim,
\newblock Nat. Commun. {\bf 4}, 2059 (2013).

\bibitem{skrzypczyk2014work}
P.~Skrzypczyk, A.~J. Short, and S.~Popescu,
\newblock Nat. Commun. {\bf 5}, 4185 (2014).

\bibitem{hardal2015superradiant}
A.~{\"U}.~C. Hardal and {\"O}.~E. M{\"u}stecapl{\i}o{\u{g}}lu,
\newblock Sci. Rep. {\bf 5}, 12953 (2015).

\bibitem{clos2016time}
G.~Clos, D.~Porras, U.~Warring, and T.~Schaetz,
\newblock Phys. Rev. Lett. {\bf 117}, 170401 (2016).

\bibitem{puglisi2017temperature}
A.~Puglisi, A.~Sarracino, and A.~Vulpiani,
\newblock Phys. Rep. {\bf 709}, 1 (2017).

\bibitem{marcantoni2017entropy}
S.~Marcantoni, S.~Alipour, F.~Benatti, R.~Floreanini, and A.~T. Rezakhani,
\newblock Sci. Rep. {\bf 7}, 12447 (2017).

\bibitem{monsel2018autonomous}
J.~Monsel, C.~Elouard, and A.~Auff{\`e}ves,
\newblock npj Quantum Inf. {\bf 4}, 59 (2018).

\bibitem{bialas2019quantum}
P.~Bialas, J.~Spiechowicz, and J.~{\L}uczka,
\newblock J. Phys. A: Math. Theor. {\bf 52}, 15LT01 (2019).

\bibitem{engquist1981definition}
H.-L. Engquist and P.~W. Anderson,
\newblock Phys. Rev. B {\bf 24}, 1151 (1981).

\bibitem{bergfield2013probing}
J.~P. Bergfield, S.~M. Story, R.~C. Stafford, and C.~A. Stafford,
\newblock ACS Nano {\bf 7}, 4429 (2013).

\bibitem{Bergfield2014Thermoelectric}
J.~P. Bergfield and C.~A. Stafford,
\newblock Phys. Rev. B {\bf 90}, 235438 (2014).

\bibitem{shastry2020scanning}
A.~Shastry, S.~Inui, and C.~A. Stafford,
\newblock Phys. Rev. Applied {\bf 13}, 024065 (2020).

\bibitem{bergfield2015tunable}
J.~P. Bergfield, M.~A. Ratner, C.~A. Stafford, and M.~{Di Ventra},
\newblock Phys. Rev. B {\bf 91}, 125407 (2015).

\bibitem{shastry2016temperature}
A.~Shastry and C.~A. Stafford,
\newblock Phys. Rev. B {\bf 94}, 155433 (2016).

\bibitem{Meair2014Local}
J.~Meair, J.~P. Bergfield, C.~A. Stafford, and P.~Jacquod,
\newblock Phys. Rev. B {\bf 90}, 035407 (2014).

\bibitem{Shastry2015Cold}
A.~Shastry and C.~A. Stafford,
\newblock Phys. Rev. B {\bf 92}, 245417 (2015).

\bibitem{Stafford2016Local}
C.~A. Stafford,
\newblock Phys. Rev. B {\bf 93}, 245403 (2016).

\bibitem{Stafford2017Local}
C.~A. Stafford and A.~Shastry,
\newblock J. Chem. Phys. {\bf 146}, 092324 (2017).

\bibitem{bevan2014first}
K.~H. Bevan,
\newblock Nanotechnology {\bf 25}, 415701 (2014).

\bibitem{morr2016crossover}
D.~K. Morr,
\newblock Contemp. Phys. {\bf 57}, 19 (2016).

\bibitem{morr2017scanning}
D.~K. Morr,
\newblock Phys. Rev. B {\bf 95}, 195162 (2017).

\bibitem{caso2010local}
A.~Caso, L.~Arrachea, and G.~S. Lozano,
\newblock Phys. Rev. B {\bf 81}, 041301(R) (2010).

\bibitem{caso2011local}
A.~Caso, L.~Arrachea, and G.~S. Lozano,
\newblock Phys. Rev. B {\bf 83}, 165419 (2011).

\bibitem{caso2012defining}
A.~Caso, L.~Arrachea, and G.~S. Lozano,
\newblock Eur. Phys. J. B {\bf 85}, 266 (2012).

\bibitem{dubi2009thermoelectric}
Y.~Dubi and M.~{Di Ventra},
\newblock Nano Lett. {\bf 9}, 97 (2009).

\bibitem{michel2003fourier}
M.~Michel, M.~Hartmann, J.~Gemmer, and G.~Mahler,
\newblock Eur. Phys. J. B {\bf 34}, 325 (2003).

\bibitem{dhar2008heat}
A.~Dhar,
\newblock Adv. Phys. {\bf 57}, 457 (2008).

\bibitem{roy2008crossover}
D.~Roy,
\newblock Phys. Rev. E {\bf 77}, 062102 (2008).

\bibitem{yang2010violation}
N.~Yang, G.~Zhang, and B.~Li,
\newblock Nano Today {\bf 5}, 85 (2010).

\bibitem{ye2016thermodynamic}
L.~Z. Ye, X.~Zheng, Y.~J. Yan, and M.~{Di Ventra},
\newblock Phys. Rev. B {\bf 94}, 245105 (2016).

\bibitem{crossno2016observation}
J.~Crossno, J.~K. Shi, K.~Wang, X.~Liu, A.~Harzheim, A.~Lucas, S.~Sachdev,
  P.~Kim, T.~Taniguchi, K.~Watanabe, T.~A. Ohki, and K.~C. Fong,
\newblock Science {\bf 351}, 1058 (2016).

\bibitem{Ye2014Thermopower}
L.~Z. Ye, D.~Hou, R.~Wang, D.~Cao, X.~Zheng, and Y.~J. Yan,
\newblock Phys. Rev. B {\bf 90}, 165116 (2014).

\bibitem{Ye2015local}
L.~Z. Ye, D.~Hou, X.~Zheng, Y.~J. Yan, and M.~{Di Ventra},
\newblock Phys. Rev. B {\bf 91}, 205106 (2015).

\bibitem{anderson1961localized}
P.~W. Anderson,
\newblock Phys. Rev. {\bf 124}, 41 (1961).

\bibitem{Tanimura1989Time}
Y.~Tanimura and R.~Kubo,
\newblock J. Phys. Soc. Jpn. {\bf 58}, 101 (1989).

\bibitem{jin2008exact}
J.~Jin, X.~Zheng, and Y.~J. Yan,
\newblock J. Chem. Phys. {\bf 128}, 234703 (2008).

\bibitem{li2012hierarchical}
Z.~H. Li, N.~H. Tong, X.~Zheng, D.~Hou, J.~H. Wei, J.~Hu, and Y.~J. Yan,
\newblock Phys. Rev. Lett. {\bf 109}, 266403 (2012).

\bibitem{Cui2019Highly}
L.~Cui, H.-D. Zhang, X.~Zheng, R.-X. Xu, and Y.~J. Yan,
\newblock J. Chem. Phys. {\bf 151}, 024110 (2019).

\bibitem{Zhang2020Hierarchical}
H.-D. Zhang, L.~Cui, H.~Gong, R.-X. Xu, X.~Zheng, and Y.~J. Yan,
\newblock J. Chem. Phys. {\bf 152}, 064107 (2020).

\bibitem{Tanimura2020Numerically}
Y.~Tanimura,
\newblock J. Chem. Phys. {\bf 153}, 020901 (2020).

\bibitem{Ye16608}
L.~Z. Ye, X.~Wang, D.~Hou, R.-X. Xu, X.~Zheng, and Y.~J. Yan,
\newblock WIREs Comput. Mol. Sci. {\bf 6}, 608 (2016).

\bibitem{Han2018On}
L.~Han, H.-D. Zhang, X.~Zheng, and Y.~J. Yan,
\newblock J. Chem. Phys. {\bf 148}, 234108 (2018).

\bibitem{Zheng2009Numerical}
X.~Zheng, J.~Jin, S.~Welack, M.~Luo, and Y.~J. Yan,
\newblock J. Chem. Phys. {\bf 130}, 164708 (2009).

\bibitem{zheng2013kondo}
X.~Zheng, Y.~J. Yan, and M.~{Di Ventra},
\newblock Phys. Rev. Lett. {\bf 111}, 086601 (2013).

\bibitem{Wang2013Hierarchical}
S.~Wang, X.~Zheng, J.~Jin, and Y.~J. Yan,
\newblock Phys. Rev. B {\bf 88}, 035129 (2013).

\bibitem{Hou2015Improving}
D.~Hou, S.~Wang, R.~Wang, L.~Z. Ye, R.-X. Xu, X.~Zheng, and Y.~J. Yan,
\newblock J. Chem. Phys. {\bf 142}, 104112 (2015).

\bibitem{Wang2018Precise}
X.~Wang, L.~Yang, L.~Z. Ye, X.~Zheng, and Y.~J. Yan,
\newblock J. Phys. Chem. Lett. {\bf 9}, 2418 (2018).

\bibitem{li2020molecular}
X.~Li, L.~Zhu, B.~Li, J.~Li, P.~Gao, L.~Yang, A.~Zhao, Y.~Luo, J.~Hou,
  X.~Zheng, B.~Wang, and J.~Yang,
\newblock Nat. Commun. {\bf 11}, 2566 (2020).

\bibitem{Yigal1993Low}
Y.~Meir, N.~S. Wingreen, and P.~A. Lee,
\newblock Phys. Rev. Lett. {\bf 70}, 2601 (1993).

\bibitem{Don0211747}
B.~Dong and X.~L. Lei,
\newblock J. Phys.: Condens. Matter {\bf 14}, 11747 (2002).

\bibitem{Dub09115415}
Y.~Dubi and M.~{Di Ventra},
\newblock Phys. Rev. B {\bf 79}, 115415 (2009).

\bibitem{Dub09042101}
Y.~Dubi and M.~{Di Ventra},
\newblock Phys. Rev. E {\bf 79}, 042101 (2009).

\bibitem{mills1998scanning}
G.~Mills, H.~Zhou, A.~Midha, L.~Donaldson, and J.~Weaver,
\newblock Appl. Phys. Lett. {\bf 72}, 2900 (1998).

\bibitem{luo1996nanofabrication}
K.~Luo, Z.~Shi, J.~Lai, and A.~Majumdar,
\newblock Appl. Phys. Lett. {\bf 68}, 325 (1996).

\bibitem{zhang2011batch}
Y.~Zhang, P.~S. Dobson, and J.~M.~R. Weaver,
\newblock Microelectron. Eng. {\bf 88}, 2435 (2011).

\bibitem{pylkki1994scanning}
R.~J. Pylkki, P.~J. Moyer, and P.~E. West,
\newblock Japan. J. Appl. Phys. {\bf 33}, 3785 (1994).

\bibitem{nakabeppu1995scanning}
O.~Nakabeppu, M.~Chandrachood, Y.~Wu, J.~Lai, and A.~Majumdar,
\newblock Appl. Phys. Lett. {\bf 66}, 694 (1995).

\bibitem{saidi2009scanning}
E.~Sa{\"\i}di, B.~Samson, L.~Aigouy, S.~Volz, P.~L{\"o}w, C.~Bergaud, and
  M.~Mortier,
\newblock Nanotechnology {\bf 20}, 115703 (2009).

\bibitem{ludovico2014dynamical}
M.~F. Ludovico, J.~S. Lim, M.~Moskalets, L.~Arrachea, and D.~S{\'a}nchez,
\newblock Phys. Rev. B {\bf 89}, 161306 (2014).

\bibitem{ludovico2018probing}
M.~F. Ludovico, L.~Arrachea, M.~Moskalets, and D.~S{\'a}nchez,
\newblock Phys. Rev. B {\bf 97}, 041416 (2018).

\bibitem{Son17064308}
L.~Song and Q.~Shi,
\newblock Phys. Rev. B {\bf 95}, 064308 (2017).

\end{thebibliography}

\end{document}